\documentclass[fleqn,usenatbib]{mnras}

\usepackage{newtxtext,newtxmath}

\usepackage[T1]{fontenc}

\DeclareRobustCommand{\VAN}[3]{#2}
\let\VANthebibliography\thebibliography
\def\thebibliography{\DeclareRobustCommand{\VAN}[3]{##3}\VANthebibliography}

\usepackage{graphicx}	
\usepackage{amsmath}	

\title[QSO identification from the slitless spectra]{Quasar identifications from the slitless spectra: a test from 3D-HST}

\author[Pang et al.]{
Yuxuan Pang,$^{1,2}$\thanks{E-mail: pyx\_s2@pku.edu.cn}
Xue-Bing Wu,$^{1,2}$\thanks{E-mail: wuxb@pku.edu.cn}
Yuming Fu,$^{3,4}$
Rui Zhu,$^{1,2}$
Tao Ji,$^{5}$
Qinchun Ma,$^{1,2}$
and Xiaotong Feng$^{1,2}$
\\
$^{1}$Department of Astronomy, School of Physics, Peking University, Beijing 100871, People's Republic of China\\
$^{2}$Kavli Institute for Astronomy and Astrophysics, Peking University, Beijing 100871, People's Republic of China\\
$^{3}$Leiden Observatory, Leiden University, P.O. Box 9513, NL-2300 RA Leiden, The Netherlands\\
$^{4}$Kapteyn Astronomical Institute, University of Groningen, P.O. Box 800, NL-9700 AV Groningen, The Netherlands\\
$^{5}$Department of Physics, The University of Hong Kong, Pokfulam Road,Hong Kong
}

\date{Accepted XXX. Received YYY; in original form ZZZ}

\pubyear{\the\year{}}

\begin{document}
\label{firstpage}
\pagerange{\pageref{firstpage}--\pageref{lastpage}}
\maketitle

\begin{abstract}
Slitless spectroscopy is a traditional method for selecting quasars. In this paper, we develop a procedure for selecting quasars (QSOs) using the 3D-HST G141 slitless spectra. We initially identify over 6,000 sources with emission lines broader than those typically found in emission line galaxies (ELGs) by analyzing the 1D spectra. These ``broad'' emission lines may originate from actual QSO broad lines ($\rm FWHM\geq1200~\rm km/s$) or the convolved narrow lines ($\rm FWHM = 200\sim 300\rm km/s$) in ELGs with effective radii $\geq$0.3" (2.5Kpc at z=1). We then propose a criterion based on the reliability of the QSO component in the forward modeling results. Using the known QSOs, ELGs, and simulation spectra, we find that our criterion successfully selects about 90\% of known QSOs with H$\alpha$ or H$\beta$ line detection and point-like structures, with an ELG contamination rate of about 5\%. We apply this method to emission line sources without significant contamination and select 19 QSO candidates at redshift $z=0.12-1.56$. 12 of these candidates have Chandra X-ray detections. This sample covers a broader range of the rest-frame UV colors and has redder optical slopes compared to the SDSS QSOs, yet it is more likely to be composed of normal QSOs rather than little red dots. Through spectral analysis, the estimated black hole masses of the sample are $10^{6.9}-10^{8.3} M_{\odot}$. Our new candidates improve the completeness of the QSO sample at $z=0.8-1.6$ in the 3D-HST field. The proposed method will also be helpful for QSO selections via slitless spectroscopy in Euclid and the Chinese Space Station Telescope.
\end{abstract}

\begin{keywords}
Active galactic nuclei (16) -- Galactic and extragalactic
astronomy (563) -- Quasars (1319) -- HST(761) -- Infrared spectroscopy(2285) -- Supermassive black holes (1663)
\end{keywords}



\section{Introduction} \label{sec:intro}
Since the first identifications of quasars (QSOs) in the 1960s \citep{1963Natur.197.1040S}, astronomers have developed plenty of selection methods for QSOs. Among these methods, one classical way is to detect their broad emission lines from the low-resolution slitless spectra. The advantage of this method is its high efficiency since, in principle, all the slitless spectra of objects in the same field of view can be obtained through one single exposure. Moreover, the selection of slitless spectroscopy is color-independent because only their emission line information is considered. However, the slitless spectroscopy technique also has several shortcomings, such as the nearby source contamination and low spectral resolution.

Various surveys from 1970s to 1990s using slitless spectroscopy successfully identified hundreds of quasars, demonstrating high recovery rates of such a method. In 1976-1979, using the prism data from the Curtis Schmidt survey in the high galactic latitude area ($|b|>40^{\circ}$), over one hundred QSO candidates were selected by their broad Lyman-$\alpha$ emission lines \citep{1976ApJ...210..267O, 1980ApJS...42..333O}, 90\% of which have been confirmed as QSOs. \cite{1982ApJ...261...18V} extended the Curtis Schmidt survey to 10 more selected areas and selected 246 QSO candidates. In the 1990s, the Palomar Transit Grism Survey (PTGS) identified 928 emission-line objects and 90 QSOs through the detections of emission lines including Ly$\alpha$+NV, CIV, CIII, MgII, H$\beta$+[OIII], and H$\alpha$ +[SII] in the wavelength range of 4400 to 7500\r{A} \citep{1994AJ....107.1245S}. The Hamburg Quasar Survey (HQS) covers about 13,600 ${\rm deg}^2$ of the northern sky ($\delta>0^{\circ}$) at Galactic latitudes $|b|>20^{\circ}$ and find about 400 QSOs by slitless survey \citep{1995A&AS..111..195H}. 
It was demonstrated that for quasars with $0.1<z<3.2$ and $B<17$, 85\% have been recovered by slitless survey \citep{1999A&AS..134..483H}. In 1990, ESO conducted another survey in the southern hemisphere called the Hamburg/ESO survey, which confirmed 400 QSOs \citep{1996A&AS..115..227W}. Using the two surveys' results, \cite{2000A&A...358...77W} analyzed the incompleteness of other color-selected QSO surveys, such as the Palomar-Green(PG) quasar survey \citep{1986ApJS...61..305G}. The basic pipeline of these surveys is to identify strong emission lines either automatically or visually (some surveys also use point-source criteria) to select QSO candidates, followed by long-slit spectroscopy observations to exclude the emission-line galaxies.

After the Sloan Digital Sky Survey (SDSS) started, QSO samples were dominated by the optical selection method in all redshift ranges. The SDSS sixteen data release contains 750,414 QSOs that have been spectroscopically identified from SDSS-I to SDSS-IV \citep[][]{2020ApJS..250....8L}. Recently, the Dark Energy Spectroscopic Instrument (DESI) data release 1 also included about 1.6 million QSOs \citep[][]{2023arXiv230606308D, 2025arXiv250314745D}.

Meanwhile, many space telescopes carry on the grism observation mode. Many slitlless spectroscopic surveys have been made with the HST (HST) to study the properties of galaxies, including the Grism Advanced Camera for Surveys (ACS) Program for Extragalactic Science (GRAPES) \citep{2004ApJS..154..501P}, the HST ACS Grism Parallel Survey \citep{2005AJ....130.1324D}, Probing Evolution And Reionization Spectroscopically (PEARS) grism survey \citep{2009AJ....138.1022S,2013ApJ...772...48P}, the Wide Field Camera Three (WFC3) Infrared Spectroscopic Parallel (WISP) Survey \citep{2010ApJ...723..104A}, 3D-HST program \citep{2012ApJS..200...13B}, the Grism Lens Amplified Survey from Space (GLASS) project \citep{2015ApJ...812..114T}, the Faint Infrared Grism Survey (FIGS) \citep{2017ApJ...846...84P}, the CANDELS Ly$\alpha$ Emission at Reionization (CLEAR) survey \citep{2019ApJ...870..133E} and the 3D-Drift And SHift (3D-DASH) program \citep{2022ApJ...933..129M}. A few studies focus on QSO properties using the ACS and WFC3 grism observation. For example, using the Palomar Transit and Hamburg slitless QSO sample, \cite{2001A&A...371...37R} find a new population of QSOs that are underluminous in X-rays. \cite{2007AAS...211.4605G} use the optical counterparts in the Chandra deepest field to find AGN features in the PEARS slitless spectra. \cite{2009ApJ...690.1181S} analyzed the helium reionization by the HST UV prism observations. The Galaxy Evolution Explorer (GALEX) also provides the grism observation in the UV band. \cite{2010ApJ...711..928C} constructed a sample of 139 Ly$\alpha$ emission line-selected sources and found that nearly all of the most luminous sources are AGNs. Based on the sample of \cite{2010ApJ...711..928C}, \cite{2010ApJ...718.1235B} constrained the QSO luminosity function and the shape of the ionizing background. Although the field of view for a single exposure in HST slitless spectra is relatively small, we are still able to select a representative sample of quasars by the grism observations as the surveyed sky area increases.

In recent years, several space telescopes with slitless spectroscopic capability have been launched or planned. Ongoing JWST programs such as the Next Generation Deep Extragalactic Exploratory Public (NGDEEP) survey, the JWST Advanced Deep Extragalactic Survey (JADES), A SPectroscopic survey of biased halos In the Reionization Era (ASPIRED), and the Cosmic Evolution Early Release Science (CEERS) survey have already demonstrated the potential of grism observations with NIRCam and NIRISS for identifying faint, high-redshift AGNs and characterizing distant galaxies properties \citep{2024ApJ...965L...6B, 2023ApJ...954L...4K, 2024A&A...691A.145M, 2023ApJ...951L...5Y}. Other telescopes, including the Euclid mission \citep{2017SPIE10563E..4WR, 2019BAAS...51c.413R}, the Chinese Space Station Telescope (CSST) \citep{2011SSPMA...41...1441, 2021ChSBu...66...1290}, and the Nancy Grace Roman Space Telescope, will conduct slitless spectroscopic surveys in large sky areas. These surveys are expected to build extensive QSO samples spanning various redshifts, reducing the color-selection biases \citep{2012MNRAS.420.1764R, 2025ApJ...980..223P}.

In this paper, we present a method for selecting QSOs from its H$\alpha$ and H$\beta$ emission lines via the 3D-HST G141 spectra. Our selection process involves two key steps. First, we identify sources with ``broad'' emission lines in the 1D spectra, following similar methods used in previous surveys. Then the more critical next step is to distinguish QSOs from emission line galaxies (ELGs) using reference images and 2D grism data. This step employs forward modeling to determine whether a source's broad emission lines originate from QSOs or are artifacts caused by extended galaxy structures. The structure of the paper is as follows: In Section \ref{sec:DATA}, we describe the data we used from the 3D-HST project, the data reduction, extraction, and forward modeling method for the grism data and the external catalog used for testing the selection criteria for known QSOs and ELGs. Section \ref{sec:separation} outlines the procedure and results for selecting QSOs in the 3D-HST G141 data, including broad emission lines identification and a specific method to distinguish QSOs from ELGs through forward modeling, identifying 19 new QSO candidates. We present the photometric properties and the contribution of new QSO candidates to the QSO luminosity function (LF) in Section \ref{sec: can_properties}. Section \ref{sec:discuss} discusses future perspectives on enhancing QSOs selection accuracy and reducing host galaxy contamination. Finally, we summarize our findings in Section \ref{sec:summary}. In this work, we adopt a flat $\Lambda$CDM cosmology with parameters $\Omega_m$ = 0.30, $\Omega_{\Lambda}$  = 0.7 and $h_0$ = 70 km $\text{s}^{-1}$ $\text{Mpc}^{-1}$.

\section{Data and Method} \label{sec:DATA}
\subsection{3D-HST project} \label{subsec:3D-HST}
The 3D-HST project is an HST treasury program providing WFC3/IR primary and ACS/WFC parallel imaging as well as grism spectroscopy over $\sim$625 arcmin$^2$ of well-studied extragalactic fields. 
3D-HST focuses on five deep fields observed by the ``Cosmic Assembly Near-infrared Deep Extragalactic Legacy Survey'' \citep[CANDELS,][]{2011ApJS..197...36K}, which are called All-Wavelength Extended Groth Strip International Survey \citep[AEGIS,][]{Davis_2007}, Cosmic Evolution Survey \citep[COSMOS,][]{Koekemoer_2007}, Great Observatories Origins Deep Survey \citep[GOODS, ][]{Giavalisco_2004} (including GOODS-North and GOODS-South), and Hubble Ultra Deep Field \citep[UDF,][]{Beckwith_2006} south, respectively. The primary data of 3D-HST was allocated 248 orbits of HST time during Cycles 18 and 19 except for the GOODS-N area, which had already been observed with WFC3/G141 before 3D-HST in a Cycle 17 program. 3D-HST also provided a photometric catalog that cross-matched with the previous deep surveys in these areas \citep{2016ApJS..225...27M}.

The 3D-HST project provides a comprehensive G141 spectra dataset in observed in 10750 to 17000\r{A} at resolution about 130, achieving a continuum signal-to-noise ratio of approximately 5 per resolution element at $H_{140}\sim23.1$ \citep{2012ApJS..200...13B}. These data have been utilized in studies of galaxy evolution and star formation \citep[e.g.,][]{2016ApJ...819..129O, 2017ApJ...850..208W}. The deep fields targeted by 3D-HST are well-documented, with existing catalogs of various objects, including QSOs and ELGs, making them ideal databases for studying QSO classification using the G141 grism spectra. Furthermore, the QSO catalog in these deep fields is currently still incomplete, providing an opportunity to discover new QSOs through a new selection method to check the completeness. 

Therefore, we selected the 3D-HST G141 grism data for constructing the QSO selection database. The data were obtained in two ways. First, we utilized the high-level science products from the 3D-HST v4.1.5 release\footnote{\url{https://archive.stsci.edu/prepds/3d-hst/}} \citep{2016ApJS..225...27M}, including 1D and 2D extracted grism spectra and multi-band photometry catalogs with emission-line information. Their slitless spectra were employed to identify sources exhibiting the "broad" emission-line features in their 1D spectra and exclude sources with severe contamination. The photometric catalogs are used to compute photometric redshifts (photo-$z$) and verify that known ELGs display emission lines in the G141 slitless spectra. Additionally, we retrieved the raw 3D-HST F140W images and G141 grism data using the \verb|mastquery.query| and \verb|astroquery.mast.Observations| modules. These raw datasets are reprocessed through data reduction and a forward modeling process.

\subsection{Grism Data Reduction} \label{subsec: grizli_intro}
We process the F140 image \& G141 data set from 3D-HST using Grizli. Grizli\footnote{\url{https://github.com/gbrammer/grizli}} \citep[the grism redshift and line analysis software, version 1.3.2,][]{2019ascl.soft05001B} performs complete end-to-end processing of the HST imaging and the slitless spectroscopic data sets by the object-based method. In this paper, we used the Grizli software to do the data reduction, astrometric alignment, contamination modeling, spectra extraction, and forward modeling.

The raw F140W/G141 data downloaded in Section \ref{subsec:3D-HST} from the MAST archive are reprocessed using the \texttt{calwf3} pipeline, with corrections for variable sky backgrounds as described by \cite{2016wfc..rept...16B}. Cosmic rays and unflagged hot pixels are identified using the AstroDrizzle software \citep{2012drzp.book.....G}. Following the approach of \cite{2016ApJS..225...27M}, the grism exposures are flat-fielded using the F140W calibration images for the G141 grisms, and the sky subtraction is performed using the “Master Sky” from \cite{2015wfc..rept...17B}. Astrometric corrections are applied by aligning the processed data to the deeper F140W HST mosaic galaxy catalog from the 3D-HST survey \citep{2014ApJS..214...24S}.

A contamination model for each 3D-HST pointing is created using a forward model based on the HST F140W full-field mosaic. The contamination model is iteratively refined, starting with a template spectrum based on a flat spectrum generated from the F140W photometry, which is assigned to sources brighter than F140W = 25 AB mag. In subsequent iterations, a third-order polynomial template spectrum is generated for sources with F140W magnitudes brighter than 24 AB mag. The result is a full contamination model for each visit of each observational program.

Grizli offers multiple methods for extracting the 1D spectra. In this study, we employed the optimal extract method, which performs a sum of the observed fluxes from all pixels within a given wavelength range using inverse variance as the weight \citep{1986PASP...98..609H}. Flux calibrations are then applied based on the sensitivity curve. The extracted spectra are used for the 1D spectra fitting process in Section \ref{subsec: phy-properties} and the 1D spectra in figures of the Appendix \ref{app: can_spec}.

Besides direct 1D spectra extraction, Grizli also provides the forward modeling method to use a linear combination of templates as the source model for fitting the 2D slitless spectra (i.e., the cubic polynomial fitting consists of four templates: $1, \lambda, \lambda^2, \lambda^3$, where $\lambda$ is the wavelength). For each input template, the procedure first calculates the 2D slitless spectra for various redshifts, based on the template spectrum and reference image. Then, a linear fit is applied to the 2D observed spectrum at each pixel to obtain the best-fit template and its corresponding redshift. Grizli provides three fitting modes: non-negative least squares (\verb|scipy.optimize.nnls|), standard least squares (\verb|numpy.linalg.lstsq|), and bounded least squares (\verb|scipy.optimize.lsq_linear|). We used the bounded least squares method and limited the flux range of each emission line to $[-10^{-16}, 10^{-13}] \rm{erg/s/cm^2}$ in the forward modeling process in Section \ref{sec:separation}.

To avoid the possible emission line degeneracy and reduce the computational cost, we further refine the fitting redshift range. For sources with spectroscopic redshifts (including known QSOs and ELGs) or two emission lines, we use a redshift fitting range of 0.15 around the true redshift or redshift given by the ASERA software \citep[A Spectrum Eye Recognition Assistant,][]{2018ascl.soft04001Y} results from Section \ref{subsec:criteria-broadline}. For sources with a single emission line, we perform the forward modeling based on photo-$z$ estimated by the broad-band photometry provided by \cite{2016ApJS..225...27M}. We train a Gradient Boosting Decision Tree (GBDT) model using the spectroscopic redshifts from known galaxies and QSOs, and then apply this model to the one-emission-line sources. Detailed information on the photo-$z$ estimation process is provided in Appendix \ref{app: photo-z}. We use these photo-$z$ results to define the redshift range for Grizli fitting, centering the redshift at the photo-$z$ value with an allowed shift $0.15\times(1+z_{photo})$.


\subsection{Known QSO and ELG Catalogs} \label{subsec: catalogs}
To establish and validate the forward modeling method for distinguishing QSOs and ELGs, as outlined in Section \ref{subsec: step2}, we cross-matched the 3D-HST photometric catalog with known QSO and ELG catalogs using a 1$''$ matching radius. To ensure enough signal-to-noise ratio (SNR) in the G141 grism spectra in sources with deeper spectral observations from ground-based surveys, we further restricted the cross-matched sources to those with F140W magnitudes brighter than 23 AB mag to keep consistent with the 5$\sigma$ detection limit per resolution element of the 3D-HST project. Additionally, to ensure that the H$\alpha$ or H$\beta$ emission lines fall within the G141 band, we constrained the redshifts of sources in the external catalogs to lie between 0.6 and 2.4. 

For the selection of known ELGs, some published catalogs do not distinguish AGNs and star-forming galaxies, which can complicate the calculation of the ELG contamination rates. Therefore, we selected spectroscopically classified ELGs from the DEEP2 and ZFIRE surveys (see details in subsections below). To further ensure that the sources from ground-based optical observations have emission lines in the G141 slitless spectra, we also require that the 
total flux of at least one emission line among H$\alpha$, H$\beta$, [OIII], or [SII] has SNR$\geq$3 in \cite{2016ApJS..225...27M} G141 emission-line catalog.

In total, we selected 35 known QSOs, 311 known ELGs from the DEEP2 survey, and 11 known ELGs from the ZFIRE survey to evaluate the selection method described in Section \ref{subsec: step2}. 
Using the DER\_SNR algorithm \citep{2008ASPC..394..505S}, we calculate the SNRs of the G141 grism spectra. Among the selected sources, the minimum SNR is 4.70, with a median value of 10.13.

\subsubsection{SDSS Quasar Catalog: the 16th data release}
SDSS \citep{2000AJ....120.1579Y} has mapped the high Galactic latitude northern sky and obtained imaging and spectroscopy data for millions of objects, including stars, galaxies, and quasars. The 16th data release of the SDSS Quasar Catalog \citep[SDSS DR16Q;][]{2020ApJS..250....8L} contains 750,414 quasars, covering four of the five 3D-HST fields (excluding GOODSS). There are 21 sources in common between the SDSS DR16Q catalog and the 3D-HST catalog. After applying magnitude and redshift constraints, 6 SDSS QSOs were selected. 

\subsubsection{The Million Quasars (Milliquas) Catalog}
The Million Quasars Catalog \citep[version 7.5;][]{2021yCat.7290....0F} is a compilation of quasars and quasar candidates from the literature. Milliquas includes nearly 1 million AGNs of various types, including type 1 QSOs, type 2 AGNs, high-confidence (pQSO = 99\%) photometric quasar candidates, and BL Lac objects. Since we want to use the typical type 1 QSOs to build up the selection method, we focused on sources labeled as “Q” in the Milliquas catalog. After applying magnitude and redshift limits and removing duplicates from the SDSS catalog, 29 additional quasars were selected for further analysis.

\subsubsection{Emission-Line Galaxy Catalog from the DEEP2 survey} \label{subsubsec: DEEP2}
The DEEP2 Galaxy Redshift Survey obtained redshifts for approximately 38,000 galaxies across four widely separated fields in right ascension \citep{2013ApJS..208....5N}, including the AEGIS field. The DEEP2 survey does not apply color pre-selection for emission-line galaxies in the AEGIS field, and it provides a spectroscopic classified redshift catalog. We cross-matched the DEEP2 emission-line galaxy catalog with the 3D-HST photometric catalog, applying magnitude and redshift constraints and removed sources which classify as ``AGN'' by DEEP2 survey or confirmed as AGN in previous studies. This resulted in a sample of 311 emission-line galaxies. The redshift range of DEEP2 selected ELGs are between 0.6 and 1.5.

\subsubsection{Emission-Line Galaxy Catalog from the ZFIRE Survey} \label{subsubsec: ZFIRE}
The ZFIRE (Z-FOURGE InfraRed Emission-line) survey is a large-scale spectroscopic effort aimed at studying galaxies in dense environments at redshifts greater than 1.5 \citep{2016ApJ...828...21N}. Utilizing the MOSFIRE spectrograph on the Keck I telescope, the survey obtains deep near-infrared (NIR) observations of star-forming galaxies in these high-density regions. Within the COSMOS field, the ZFIRE survey has identified over 200 emission-line galaxies, primarily around redshift 2, which is higher than the DEEP2 survey. To select a sub-sample for further analysis, we cross-matched the ZFIRE emission-line galaxy catalog with the 3D-HST photometric catalog and applied the magnitude and redshift constraints. From this cross-match, we identified 11 galaxies that exhibit H$\beta$ or [OIII] emission lines in the G141 band within the redshift range 2.1—2.4.

\subsection{SDSS spectra of QSOs and ELGs} \label{subsec: sdss-spec}
We performed slitless spectroscopic simulations using spectra from SDSS DR16 classified as QSOs, star-forming galaxies, and starburst galaxies \citep{2020ApJS..249....3A}. In order to match the rest-frame wavelength range of the G141 band in redshift 0.6–2.4 and ensure the quality of the simulated spectra and their emission line features, we downloaded spectra of 150 QSOs, 120 star-forming galaxies and 80 starburst galaxies by the following criteria:
\begin{itemize}
    \item 1. The spectra have been successfully classified (\verb|class=='QSO'| for QSOs, \verb|class==’GALAXY’ AND subclass in ('STARFORMING',| \verb|'STARBURST')| for ELGs). 
    \item 2. The spectra should have redshift in 0-0.3 (\verb|Z>0 AND Z<=0.3|) and no known problems in redshift measurement (\verb|ZWARNING == 0 OR ZWARNING == -1| for QSOs, \verb|ZWARNING == 0| for galaxies).
    \item 3. The spectra has high SNR (\verb|SN_MEDIAN_ALL > 5|).
    \item 4. Visually confirmed that the \rm{H$\alpha$} or \rm{H$\beta$} emission lines are presented in the spectra.
\end{itemize}

\section{QSO selection in 3D-HST data} \label{sec:separation}
\begin{figure}
\includegraphics[width=0.47\textwidth]{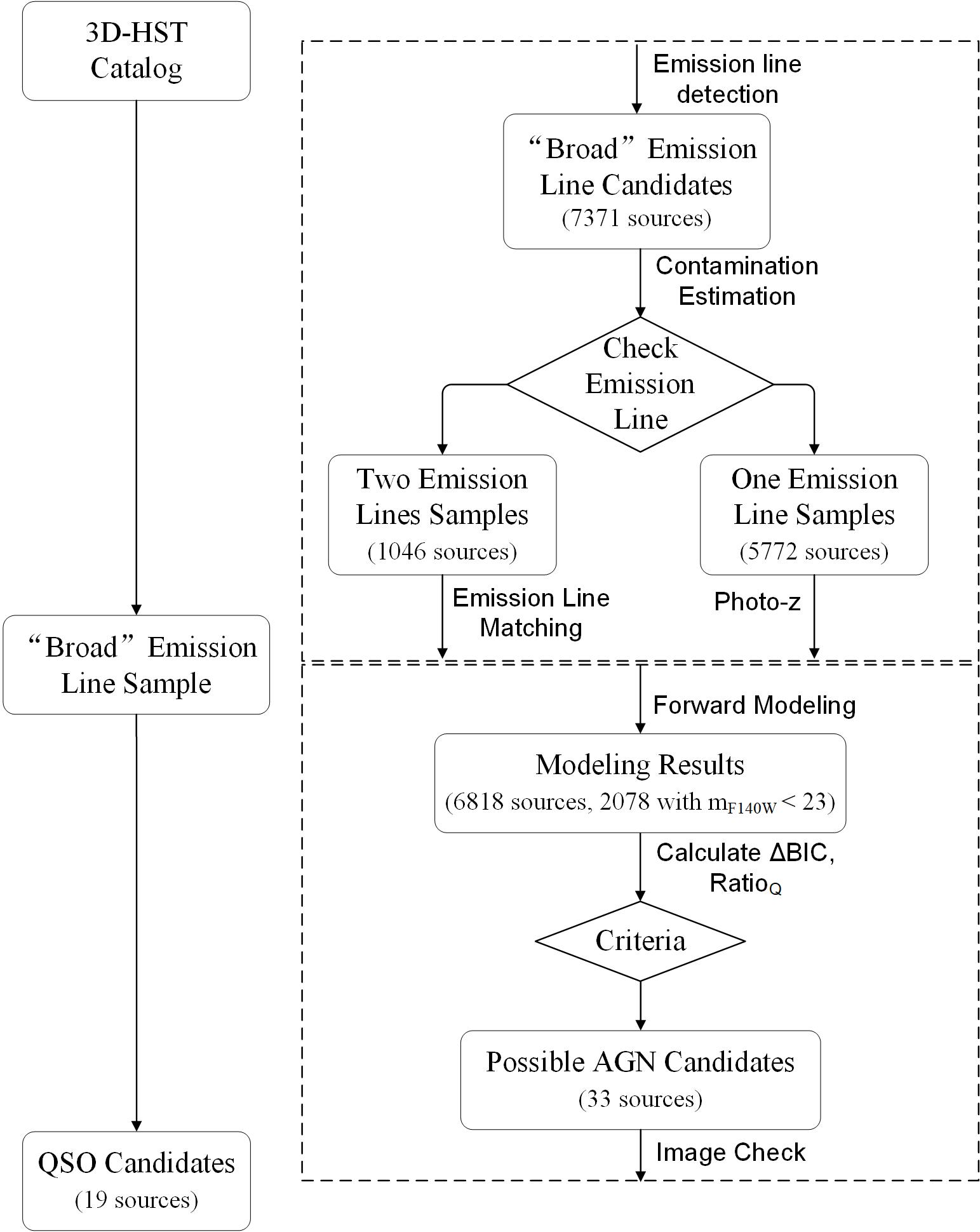}
\caption{The procedure for detecting QSOs from 3D-HST grism data. The left side of the panel illustrates the two-step process, while the right dashed boxes show the specific steps.
\label{fig: logic_graph}}
\end{figure}

In this section, we present a two-step method for selecting QSOs from the G141 grism spectra. First, in Section \ref{subsec:criteria-broadline}, we describe the details for identifying sources with ``broad'' emission lines in the 1D spectra. Then, in Section \ref{subsec: step2}, we develop a method to distinguish QSOs from ELGs based on forward modeling results derived from the known QSOs and ELGs. Figure \ref{fig: logic_graph} summarizes the overall procedure of our method.

\subsection{Step1: Broad emission line detection} \label{subsec:criteria-broadline}
\begin{figure*}
\includegraphics[width=0.95\textwidth]{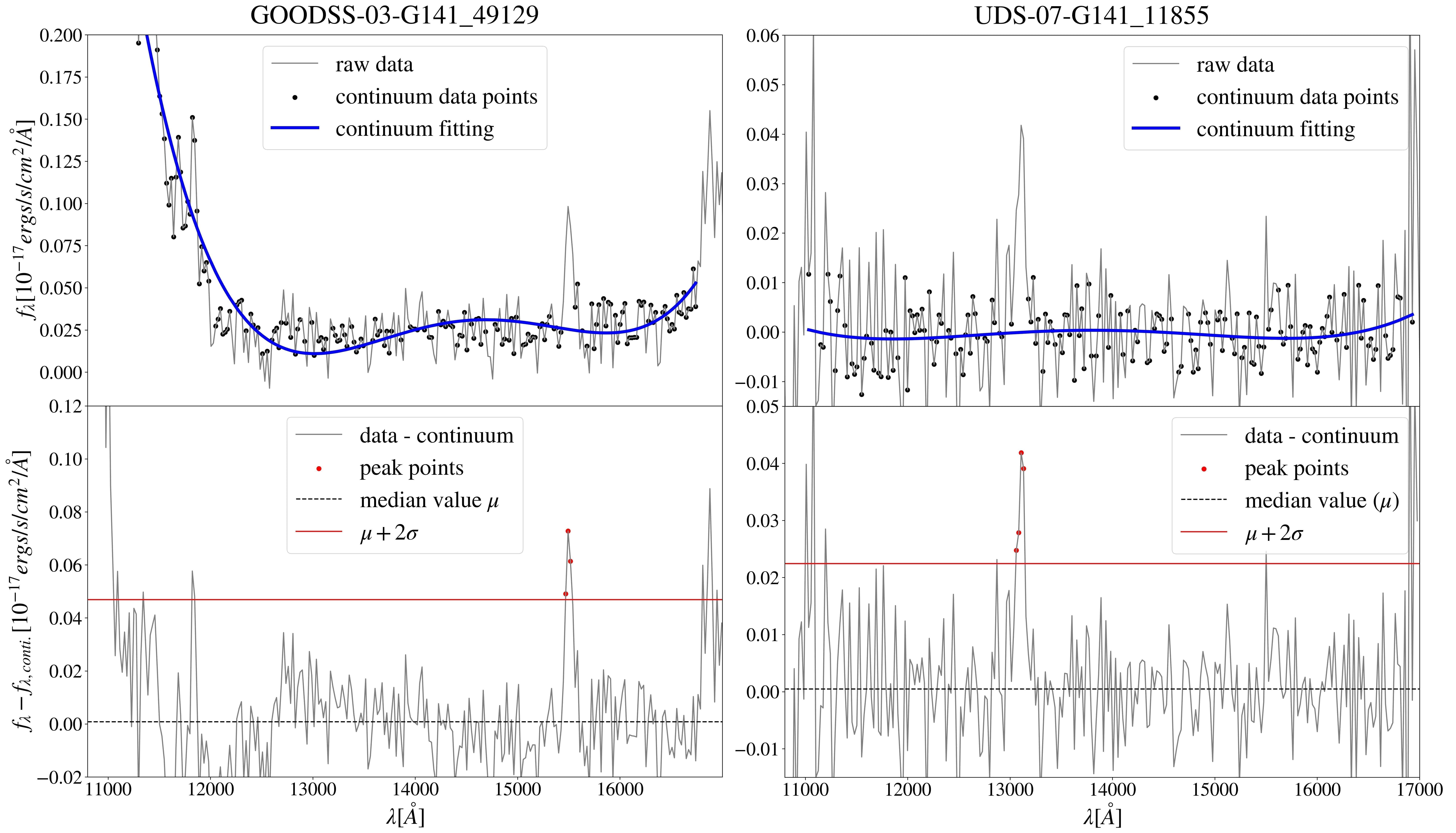}
\caption{Illustrations of the emission line detection methods in 1D spectra from the 3D-HST v4.1 data release. The left figure shows a distorted spectrum with an irregular continuum, while the right figure depicts a relatively normal emission line source. In each figure, the upper panels display the continuum fitting process: the grey lines represent the raw data, the black dots indicate the data selected for continuum fitting, and the blue lines show the quartic fit to the continuum. The lower panels present the results of our emission line detection, using the median and standard error of the signal derived from subtracting the fitted continuum from the data. The grey lines represent the signal; the black dotted lines indicate the signal's median, and the red line marks the judgment threshold, set to be larger than two standard deviations from the median. The red dots highlight the emission line peaks identified by this threshold.
\label{fig:emission_line_judge}}
\end{figure*}

The 3D-HST project provides over 200,000 G141 grism spectra, comprising $\sim$22,000 sources with the detected H$\alpha$, H$\beta$, or [O~\textsc{III}] emission lines at SNR \textgreater~3. To reduce the sample size for detailed fitting, we initially selected sources with ``broad'' emission lines, which are more likely to be QSOs. The emission-line identification procedure is outlined below, with Figure \ref{fig:emission_line_judge} demonstrating two representative examples of this detection process:

\begin{itemize}
    \item 1. Reorganizing and combining all the 1D spectra from the 3D-HST data release v4.1 for each source.
    \item 2. Detrending the continuum of the 1D spectra. To address potential fluctuations in the low-sensitivity regions at the blue and red ends of the grism and to mitigate unusual spectral shapes caused by the contamination or the convolution effects, we estimate the continuum using a quartic polynomial model. To reduce the influence of emission lines and individual outlier points on the continuum, we divide the data into four wavelength bins and combine the 20th to 80th percentile of each bin's data to fit the quartic function. To ensure sufficient data points for continuum fitting, we exclude sources with fewer than 10 (wavelength, flux) points above a 3$\sigma$ level in the slitless spectra.
    \item 3. Identifying emission line features. After subtracting the continuum from the 1D spectra, we detect signals that exceed 2$\sigma$ above the average over three consecutive data points. Such emission line signal corresponds to a full width at half maximum (FWHM) larger than 2000 km/s.
\end{itemize}

\begin{figure*}
\includegraphics[width=0.95\textwidth]{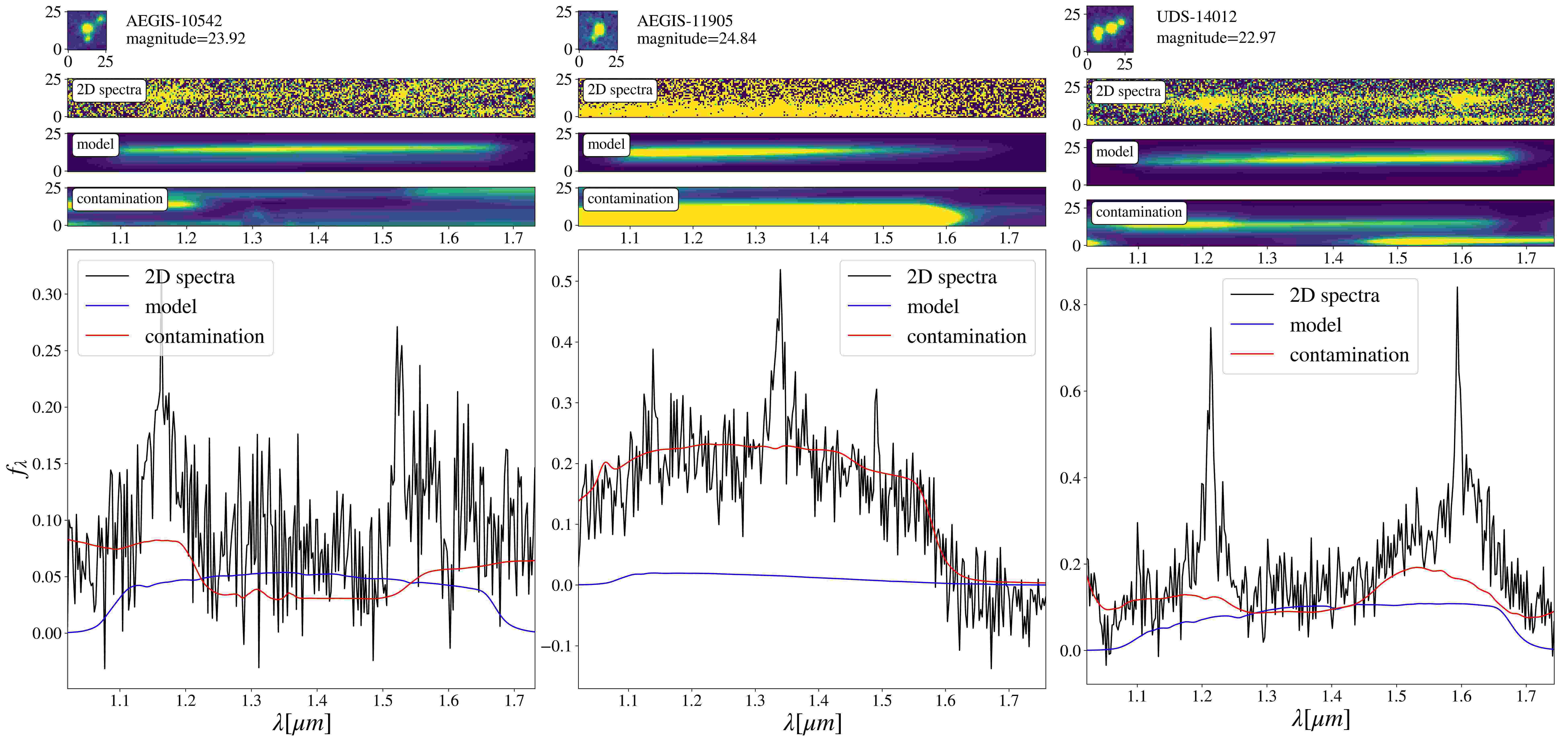}
\caption{Illustration of sources that have serious contaminations. The 2D and 1D slitless spectra are taken from the 3D-HST data release \citep{2016ApJS..225...27M}. The upper left panel in each figure gives the reference image. The following three 2D figures show the 2D slitless spectra, model, and contamination model provided by the 3D-HST data release. The lower panels compare the model and contamination shown as blue and red lines, respectively.
\label{fig:contamination_judge}}
\end{figure*}

Before further analysis, we exclude sources with contamination flux exceeding the model flux at all dispersion angles in the 3D-HST v4.1 data release. Figure \ref{fig:contamination_judge} provides examples of the contamination sources. For sources with two or more detected emission lines, we also conducted visual inspections using the ASERA software. During this process, we excluded sources with the following features:

\begin{itemize}
    \item Anomalous Line Profile: $\geq$3 distinct peaks in the H$\alpha$ complex, excluding contributions from the adjacent [S II] doublet at 6717,6731\r{A}; or $\geq$4 peaks across the H$\beta$+[O~\textsc{III}] 4959,5007\r{A} complex.
    \item Non-Physical Line Widths: emission line with FWHM > 12,000 km/s, exceeding the velocity widths of 99\% of Type 1 quasars in the SDSS and LAMOST QSO surveys \citep{2011ApJS..194...45S, 2023ApJS..265...25J}.
    \item Emission line Mismatches: at least one emission line (SNR > 3) deviates from the systemic redshift by $\Delta z > 0.015$ relative to other lines. These mismatched lines usually are the 0th-order contamination from the other sources.
\end{itemize}

\begin{figure}
\includegraphics[width=\columnwidth]{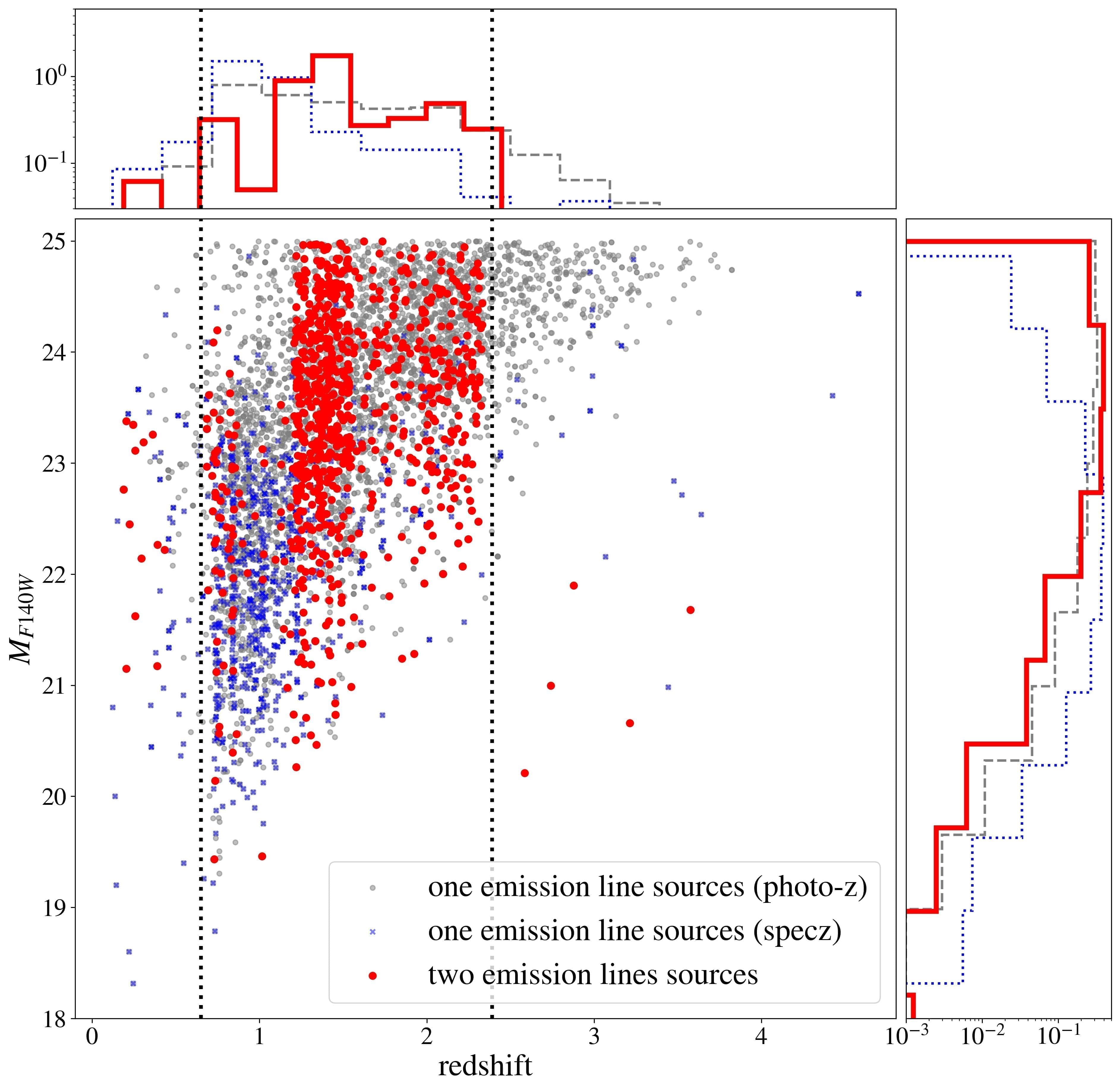}
\caption{The redshift and F140W AB magnitude distribution of sources in our one  and two emission line (red dots) samples. The ASERA visual inspection results give the redshift of two emission line sources, while for the one emission line source, the redshift is given by the previous spectroscopy results (in blue cross) or photometric redshift estimation (in grey dots). The black dotted lines show that most of our sample lies in the range of redshift 0.6-2.4, where either H$\alpha$ or H$\beta$ \& [OIII] emission line is located in the range of the G141 band. (In this figure, we applied a cutoff at 25 AB magnitude to keep consistent with 3D-HST detection limit.)
\label{fig:emission_line_properties}}
\end{figure}

Our broad emission line catalog includes 1,046 sources with two emission lines after excluding 74 sources with contamination and 86 sources with data issues, and 5,772 sources with one emission line after excluding 393 sources with contamination. Figure \ref{fig:emission_line_properties} displays our cataloged sources' F140W magnitude distribution and redshift from ASERA software or photo-$z$ estimation. Most sources are in the redshift range of 0.6 to 2.4, with H$\alpha$ and H$\beta$ emission lines within the G141 grism band. All known QSOs matched in Section \ref{subsec: catalogs} are successfully recovered by our broad-line selection pipeline and unaffected by contamination or other data issues.

\subsection{Step2: Separate QSOs from ELGs} \label{subsec: step2}
\begin{figure*}
\includegraphics[width=1.0\textwidth]{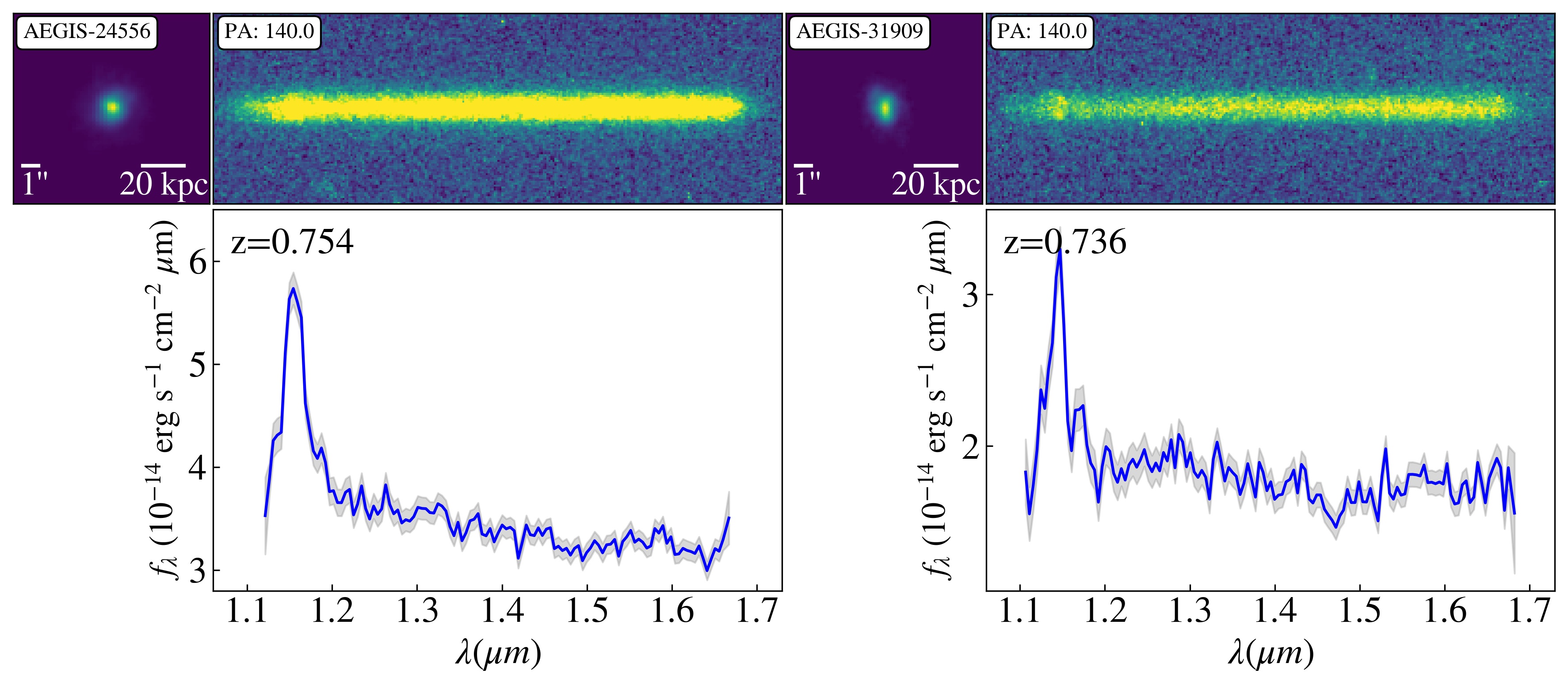}
\caption{Examples of two ELGs confirmed by the DEEP2 survey that have 1D G141 grism spectra resembling type 1 QSOs, AEGIS-24556 (left panel) and AEGIS-31909 (right panel). The top row shows the F140W image alongside the 2D G141 grism data, while the bottom row displays the extracted 1D spectra, the note on the upper left shows the source name, observation grism, PA (angle East of North) of the dispersion axis, and the redshift. Based on the DEEP2 spectroscopy redshift, the emission line in the G141 grism corresponds to the narrow \rm{H$\alpha$} line, however, the shape of the emission line appears to resemble the low-resolution \rm{H$\alpha$} lines of an AGN with a FWHM>1200km/s, due to the convolution effects.
\label{fig:mixture of ELG and QSO}}
\end{figure*}

In the previous step, we identified a large number of emission line sources using G141 slitless spectra. Due to the convolution effect of extended morphology in the images, narrow emission lines from ELGs can appear to be similar to the broad emission lines of QSOs in the slitless spectra, making many of these sources difficult to classify. Figure \ref{fig:mixture of ELG and QSO} illustrates an example of two emission line galaxies. Their DEEP2 spectra of both sources show narrow \rm{H$\beta$} emission line, as well as Balmer absorption features \citep{2013ApJS..208....5N}, but their G141 grism spectra displays a ``broad'' emission line similar to type 1 QSOs. In this section, we employ the forward modeling method using the Grizli package to establish a selection criterion for distinguishing QSOs from ELGs.

\subsubsection{QSO selection criteria} \label{subsec:grizli-criteria}
The core principle of our selection method relies on the distinct 2D emission line profiles of QSOs and ELGs. For point-like QSOs, their emission line morphology appears elongated along the dispersion direction compared to their reference images. Consequently, incorporating a QSO template with broad emission-line components yields significantly better fits in emission-line wings than those only modeled with the narrow emission-line complexes of galaxies. In contrast, galaxies’ 2D emission-line maps contain spatially resolved \rm{H~\textsc{II}} regions whose distribution diverges from their reference images, making single-template modeling insufficient to reconstruct their 2D emission-line structure. This limitation results in minimal variation in the residuals of ELG fits when a QSO template is added to the model. Leveraging this divergence, we employ two template sets to analyze 2D slitless spectra, identifying QSOs by quantifying differences in the fitting performance between two models.


The first group uses the Flexible Stellar Population Synthesis (FSPS) templates \citep{2009ApJ...699..486C, 2010ApJ...712..833C}, including emission line complexes with fixed ratios \citep[see more details in][]{2023ApJS..266...13S}. The second group expands on the templates of the first group by adding the SDSS QSO composite template \citep{2001AJ....122..549V}. The SMC extinction curve \citep{2001ApJ...548..296W}, which has been shown to effectively model the dust reddening of most quasars at low to intermediate redshifts \citep{Richards_2003,Hopkins_2004}, is applied to our templates. To enhance the flexibility of the model and mitigate degeneracies between overly red QSO templates and galaxy templates, we introduce an extinction parameter Av ranging from 0 to 0.6 which encompasses the extinction values observed in the majority of QSOs \citep{Baron_2016}. For clarity, we refer to the fitting results with and without the QSO template as the ``Q'' and ``G'' groups, respectively, in the following sections. To further illustrate the fitting results produced by Grizli, our figures follow the plotting structure in the software Grizli\_extra\footnote{\url{https://github.com/Fmajor/grizli_extras}}.

\begin{figure*}
\includegraphics[width=1.0\textwidth]{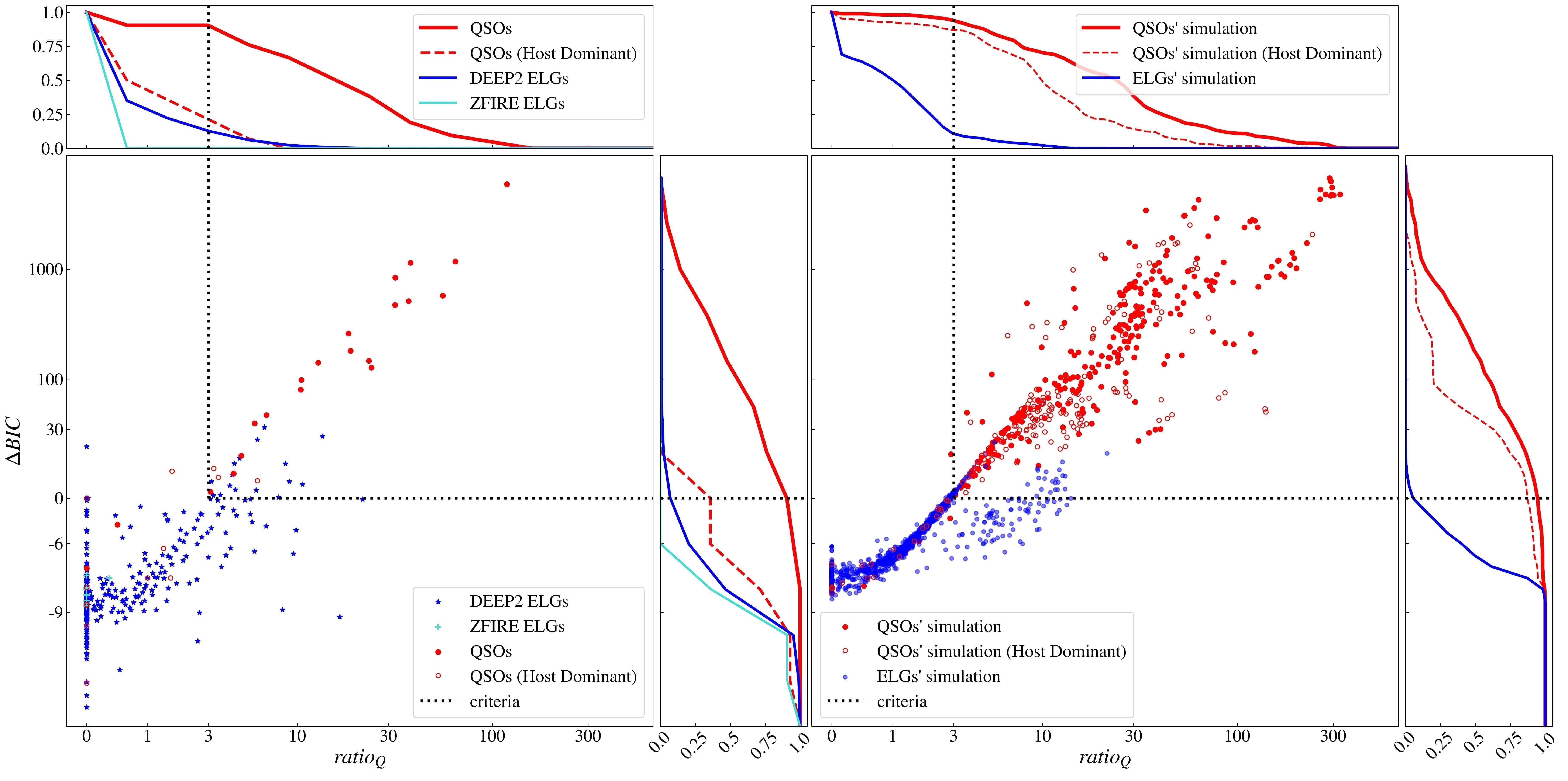}
\caption{The distribution of QSOs and ELGs in a parameter space defined by the ratio of QSO composite spectral coefficients ($ratio_{Q}$, horizontal axis) and the difference in the Bayesian information Criterion ($\Delta BIC$, vertical axis). The top and right-side insets display cumulative distribution functions, indicating the proportion of sources with parameter values above a given threshold. The black dotted lines show the selection criteria defined in Formula \ref{formula:criteria} and \ref{formula:criteria2}. \textbf{Left panel}: Results from known QSOs and ELGs in external catalogs. Solid red points (solid red lines in the cumulative panels) represent Point-Like known QSOs, while hollow red points (dashed red lines) indicate Host-Dominant known QSOs (defined in Section \ref{subsec:QSO-result}). Blue and cyan points (and corresponding lines) represent ELGs from the DEEP2 and ZFIRE surveys, respectively. \textbf{Right panel}: Results from simulated G141 grism data. Solid red points (solid red lines) correspond to simulations based on Point-Like known QSOs (Group 1 in Section \ref{subsec:simu-result}), while hollow red points (dashed red lines) are based on Host-Dominant known QSOs (Group 2). Blue points (solid blue lines) represent simulations based on known ELGs (Group 3).
\label{fig:criteria-figure}}
\end{figure*}

After the fitting, we extract the linear fit coefficients ($\rm Coeff$) and their corresponding errors ($\rm error$) from the ``Q'' group templates, as well as the best-fit $\chi^2$ values for both the ``Q'' and ``G'' groups. Since the ``Q'' group includes the additional QSO template for linear fitting, the $\chi^2$ values for all ``Q'' group fits are expected to be smaller than those for the ``G'' group. To determine whether adding the QSO template truly improves the fit, we applied the Bayesian Information Criterion (BIC) as a selection criterion \citep{1978AnSta...6..461S,1311138,2021MNRAS.501.2268W}. BIC is a statistical measure that indicates whether adding an extra fitting parameter improves the model, and it is defined as:
$$\rm{BIC}=k \rm{ln}(n)+\chi^2,$$ \label{formula:BIC}
where $n$ is the number of data points used for fitting and k is the number of templates with non-zero coefficients. Using this formula, we calculated the BIC values for the best fits of the ``Q'' and ``G'' groups. Subsequently, we calculated the Bayesian Information Criterion difference ($\Delta \rm{BIC}$) between the BIC values from the ``Q'' and ``G'' groups and the ratio ($\rm{ratio}_Q$) between the coefficient ($\rm{Coeff}_Q$) and error ($\rm{error}_Q$) of the QSO composite spectrum model. Finally, we used the $\Delta BIC$ and $\rm{ratio}_Q$ parameters to distinguish QSOs from emission-line galaxies (ELGs). The essential criteria are:
\begin{gather}
    \Delta \rm{BIC}=\rm{BIC}_{G}-\rm{BIC}_{Q}>0, \label{formula:criteria}\\
    \rm{ratio}_Q=\rm{Coeff}_{Q}/\rm{error}_{Q}>3. \label{formula:criteria2}
\end{gather} 
The first criterion ensures that adding a QSO component improves the fit in the 2D slitless spectrum. On the other hand, the second criterion provides a high confidence level for the presence of the QSO component. Figure \ref{fig:criteria-figure} shows the two-dimensional space of two parameters, with the known QSOs (located in the right upper corner) and ELGs (located in the left lower corner) are displayed in red and blue colors. 

For each source, we obtained the pure emission line (EL) maps for H$\alpha$
and [O \textsc{III}] by subtracting the continuum components from the fitted FSPS and QSO composite spectra. To further separate the partially blended [O \textsc{III}] 4959,5007\r{A} doublets, we adopted a de-blending technique similar to those employed in \cite{2015AJ....149..107J} and \cite{2017ApJ...837...89W}. The source contour is determined using the segmentation map from the F140-band reference image. This contour is then applied to the 2D grism spectra at the observed wavelengths corresponding to the [O \textsc{III}] doublet to measure the initial flux. Utilizing the theoretical [O \textsc{III}] doublet flux ratio of 2.98 \citep{2000MNRAS.312..813S} which is also consistent with SDSS DR7 QSO results \citep{2011ApJS..194...45S}, we iteratively subtract the flux contribution from the [O \textsc{III}] 4959\r{A} line until complete removal is achieved, yielding the isolated [O \textsc{III}] 5007\r{A} emission map.

\subsubsection{Known QSOs Performance} \label{subsec:QSO-result}
\begin{figure*}
\includegraphics[width=1.0\textwidth]{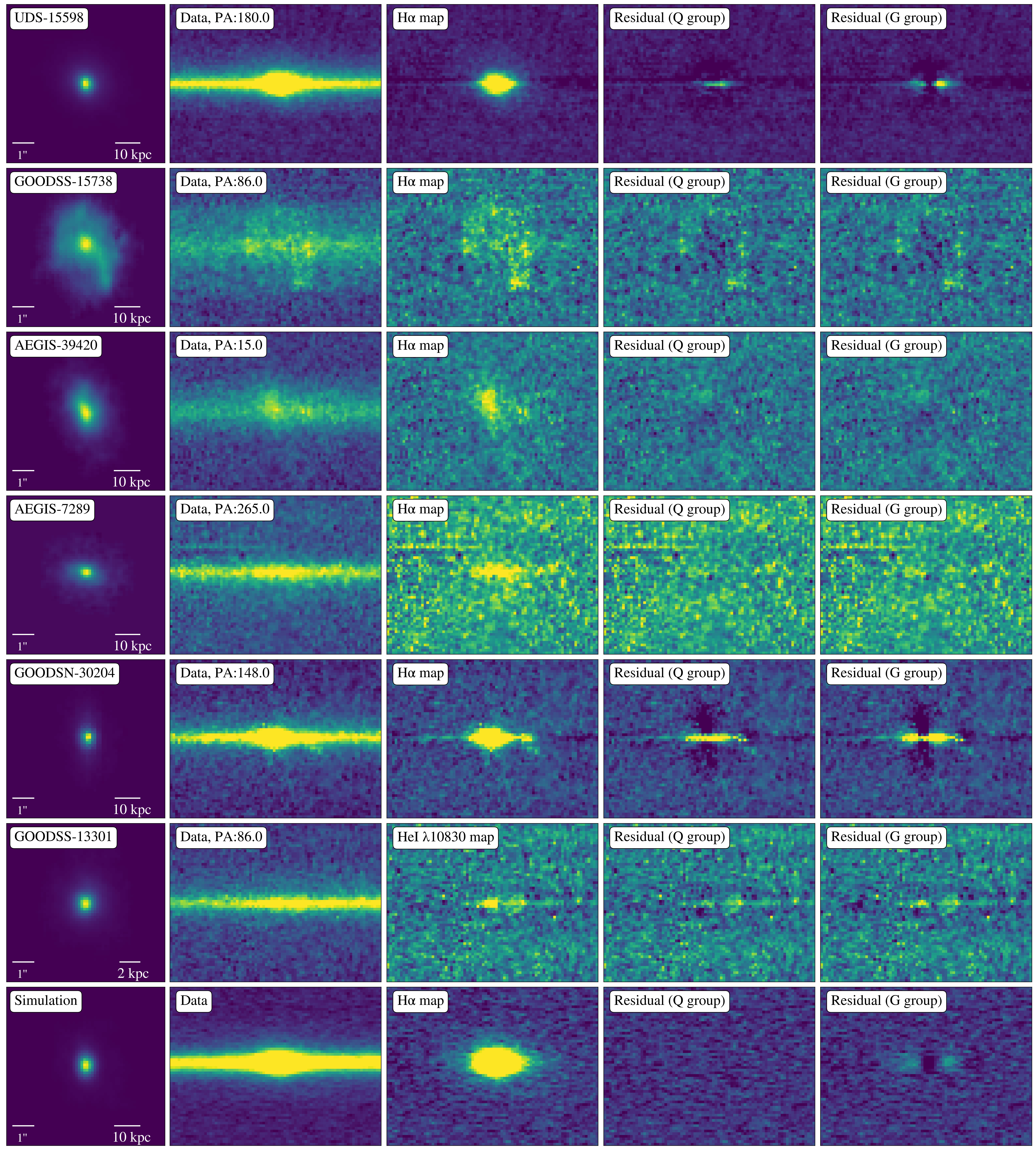}
\caption{2D emission line map and modeling results of typical objects in G141 slitless spectra. Columns from left to right: F140W reference image (the source name annotated at upper right corner; angular scale and physical scale indicated below), 2D G141 slitless spectrum (the position angle [PA; angle East of North] labeled at upper right), emission morphology after continuum subtraction (emission line name annotated at upper right corner), residual maps from the "Q" group modeling (with the QSO template), and residual maps from the "G" group modeling (without the QSO template). \textbf{Rows 1–2:} Known QSOs. UDS-15598 (Row 1) is a Point-Like QSO satisfying the criteria (Section 3.2.2), while GOODSS-15738 (Row 2) is a Host-Dominant QSO. \textbf{Rows 3–4:} Emission-line galaxies from the DEEP2 survey. AEGIS-39420 (Row 3) does not meet our criteria, whereas AEGIS-7289 (Row 4) falls into the selection region. \textbf{Row 5-6} (GOODSN-30204, GOODSS-13301): Newly identified QSO candidates. \textbf{Row 7:} Simulation results from Group 1 (Section 3.2.4). Comparison of residual maps (last two columns) reveals that for typical QSOs (Rows 1, 2, 5, 6, 7), the "Q" group modeling produces better fits in the line wings compared to the "G" group. For galaxy-dominated sources (Rows 2–3), minimal differences exist between the "Q" and "G" residuals. The contaminating ELG in Row 4 shows slightly improved "Q" group residuals near the line core, likely due to data fluctuations or localized emission-line knot structure.
\label{fig: fitting_re}}
\end{figure*}

We calculated the $\Delta$BIC and $ratio_{Q}$ values of 35 known QSOs selected in Section \ref{subsec: catalogs}. 22 QSOs satisfied the selection criteria outlined in Section \ref{subsec:grizli-criteria}, yielding a detection rate of 63\%. As shown in the first row of Figure \ref{fig: fitting_re}, the 2D emission line map and fitting results of a representative QSO meeting our criteria demonstrate the noticeable improvement in modeling the extended wings of the H$\alpha$ emission line.

\begin{figure}
\includegraphics[width=\columnwidth]{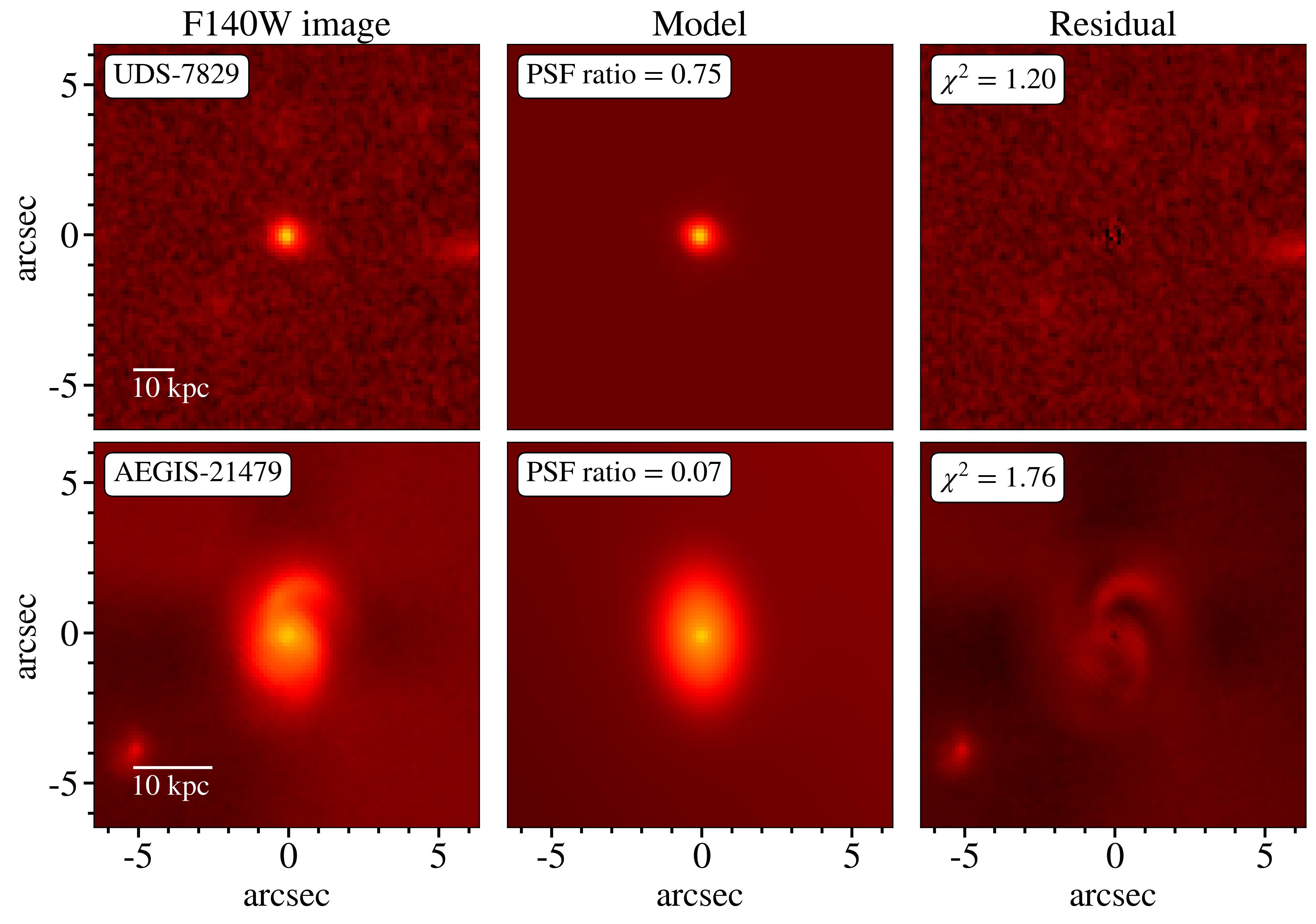}
\caption{Examples of F140W image decomposition results. Three panels from left to right show the F140W image, Galfit model, and residual. The note in each panel gives the object name, the PSF component flux ratio, and the reduced $\chi^2$ of the fitting results, respectively. The upper panel shows the results of UDS-7829 which contains a relatively high PSF component ratio, while the lower panels show the results of a host dominant source AEGIS-21479.
\label{fig:image_decom}}
\end{figure}

We further analyzed why the broad emission-line features of some QSOs cannot be effectively identified through forward modeling. We found that for certain QSOs, their 2D emission-line features are not as well-defined as the typical QSO spectrum, which usually displays a narrow, elongated broad line along the dispersion direction. Instead, these objects show clumpy features, similar to galaxies. In these cases, the presence of resolved or partially resolved star-forming ``knots'' or ``clumps'' in the host galaxy still contaminates the 2D grism spectra, particularly across the emission lines. This blurring of the broad emission-line component from the central AGN complicates the forward modeling process. To further investigate the impact of the host galaxy on the selection efficiency, we decomposed the F140W images of the 35 known QSOs into a PSF component by the pattern from \cite{2014ApJS..214...24S} and a Sérsic component. Figure \ref{fig:image_decom} provides typical examples of sources dominated by the AGN versus those dominated by the host galaxy. Although the decomposition method cannot fully resolve the detailed structures like spiral arms in quasar host galaxies (e.g., AEGIS-21479 in Figure \ref{fig:image_decom}), this approach can effectively characterize the relative flux contributions between the central AGN and host galaxy for known quasars, thereby evaluating the host galaxy's influence on our modeling. We found that using a 25\% threshold for the central PSF flux in the F140W image provides a good distinction for understanding the host galaxy's influence. For convenience, we define the Point-Like QSOs as those where the central PSF flux accounts for more than 25\% of the total flux, and the Host-Dominant QSOs as those where the host galaxy flux exceeds 75\%. Among 35 known QSOs, 21 are Point-Like, and the other 14 are Host-Dominant.

For 21 Point-Like QSOs, 19 met our selection criteria, showing a high success rate. The remaining two sources that did not meet our criteria are COSMOS-5265 and UDS-5629. COSMOS-5265 has a redshift of 1.869, with H$\beta$ and [O \textsc{III}] emission lines detected in the G141 band. This source is relatively faint compared to other QSOs in our sample (with a magnitude of approximately 22.92 in the F140W band), and its continuum shape is redder than the typical QSO power law. UDS-5629 exhibits a relatively flat continuum and a weak H$\beta$ emission line. This source has been confirmed as a QSO by the Subaru-XMM-Newton Deep Survey \citep[SXDS;][]{2015PASJ...67...82A}.

For 14 Host-Dominant QSOs, only 3 met our selection criteria. As shown in the second row of Figure \ref{fig: fitting_re}, the H$\alpha$ 2D emission line map of a typical Host-Dominant QSO reveals significant contamination from the host galaxy's star-forming regions, with limited broad emission line signal from the center AGN, therefore the ``Q'' and ``G'' groups give similar fitting results and fails to meet our selection criteria. We found that the remaining 11 sources are all from the Milliquas catalog. Six of the 11 sources originate from \cite{2019MNRAS.487.4285P}, which aimed to select faint AGNs through HST variability, and these sources may not represent typical Type 1 QSOs. The remaining five sources are part of follow-up observations for X-ray AGN selection programs in deep fields. These sources, which exhibit weaker AGN components compared to typical QSOs, require higher SNR observation and more detailed modeling to separate the AGN and host galaxy SEDs. 

Overall, the forward modeling given by Grizli proves to be effective in selecting the Point-Like known QSOs (19/21, 90\%). However, due to the limitations of single-PA observations, the selection performance for the Host-Dominant known QSOs is limited (3/14, 21\%), due to the self-contamination by the host galaxy. In total, the QSO selection success rate is 63\% (22/35).

\subsubsection{Known ELGs Performance} \label{subsec:ELG-result}
To estimate the pollution rate of ELG to QSO selection, we performed forward modeling and calculated the $\Delta$BIC and $ratio_{Q}$ values for known ELGs from DEEP2 and ZFIRE surveys.

Among 311 emission-line galaxies in the DEEP2 survey in the AEGIS field, with redshifts ranging from 0.6 to 1.5, 15 sources satisfy our QSO selection criteria, resulting in an overall contamination rate of approximately 4.8\%. This contamination estimate should be considered conservative, as the template-fitting spectral classification methodology employed by DEEP2 shows reduced reliability for low-luminosity targets. We therefore visually inspected the spectra of all emission-line galaxies from the DEEP2 survey that met our QSO selection criteria. All 15 identified sources show [O \textsc{II}] 3726, 3729\r{A} emission lines, but 6 of them exhibit prominent higher-order Balmer absorption lines, suggesting they are not typical Type 1 AGNs. The remaining sources are still challenging to conclusively determine the presence of weak AGNs at their centers.

Further analysis of the G141 grism data of these sources reveals two types of contamination. Firstly, most of the sources (11/15) are influenced by the emission-line features from the galaxy itself, especially when the nucleus and star-forming knots appear along the same axis in the grism dispersion angle (8 cases in 11 sources). As shown in the fourth row of Figure \ref{fig: fitting_re}, the modeling results of the star-forming regions may nonphysically improve by adding the QSO components when the regions shows a more extended morphology compared to the reference image. The remaining four sources are located in very close pairs. In these cases, the reference image is affected by the nearby sources, leading to incomplete emission-line modeling.

Therefore, the ELG contamination cases will show typical morphology features in the reference image and the 2D emission line map, making them removable by further visual image identification. Additional observations at multi-PAs could also help to break up the degeneracy automatically. 

%

In the COSMOS field, among the 11 emission-line galaxies from the ZFIRE survey with redshifts larger than 2, all sources were excluded from our QSO selection region.

\subsubsection{Simulation result} \label{subsec:simu-result}
In previous two sections, we presented the results of known QSOs and ELGs. To further validate the performance of the forward modeling method in an idealized scenario, we conducted additional simulations. First, we convolved the F140W reference image data of known QSOs and ELGs with their redshifted QSO/ELG spectra from SDSS spectra to generate the original 2D grism data. Then, we added noise to reflect the SNR typical of slitless spectra. The simulated 2D grism data and their corresponding F140W reference images are then used as simulated input for forward modeling. This idealized simulation removes the influence of self-contamination, enabling us to further investigate the limitations of forward modeling due to data degeneracies, errors, and intrinsic spectral effects. The last row in Figure \ref{fig: fitting_re} shows an example of the simulation and the fitting result of the H$\alpha$ 2D map of a Point-like QSO, which shows similar improvement in modeling the extended wings in ``Q'' group as the known QSO results in the first row.


As shown in the right panel of Figure \ref{fig:criteria-figure}, we performed three groups of simulations. Group 1 used the F140W image of the Point-Like known QSOs from Section \ref{subsec:QSO-result} and their corresponding redshifted QSO spectra as input, simulating 271 G141 grism spectra. The fitting results show a loss rate of about 6\%, slightly smaller than the loss rate for Point-Like known QSOs discussed in Section \ref{subsec:QSO-result}. This result confirms that a small fraction of QSOs, particularly those with significantly different spectra (e.g., narrower emission line profiles) compared to the QSO composite spectra, may be missed by the selection criteria.

Group 2 used the F140W images of the Host-Dominant known QSOs from Section \ref{subsec:QSO-result} along with their redshifted QSO spectra, simulating 193 spectra. In this case, the simulation showed a loss rate of about 13\%, much larger than Group 1 but still smaller than the loss rate for the Host-Dominant known QSOs in the real case. This illustrates that the selection failures for the Host-Dominant known QSOs in Section \ref{subsec:QSO-result} are primarily due to self-contamination from the star-forming regions in the host galaxy.

Group 3 used the F140W images of ELGs from the DEEP2 and ZFIRE surveys and redshifted star-forming/starburst spectra as input. The simulation indicated an overall contamination rate of 4\% (slightly lower than the contamination rate observed in Section \ref{subsec:ELG-result}). Notably, when the major axis of the galaxy is aligned with the dispersion axis (within 20°), the contamination rate increases to about 6\%. This result highlights that, in addition to the self-contamination caused by the distribution of star-forming regions aligned with the dispersion axis, the inherent model degeneracy due to the single dispersion direction also leads few ELG to be misidentified as QSOs.

\subsubsection{The criteria performance}
Our selection criteria, which are based on improvements in forward modeling, demonstrate high accuracy in identifying QSOs with minimal host galaxy contamination. For 35 known QSOs in our sample, our criteria successfully selected 22 sources (62.8\%). The remaining unsuccessful cases primarily involve AGNs dominated by host galaxy light in F140W bands, where contamination from the host galaxy's emission line features and insufficient SNR of the central broad emission line present challenges. For known ELG samples, our criteria effectively exclude most known ELGs, with only 15 out of 311 DEEP2 survey sources in the AEGIS field identified as potential contaminants ($\sim$4.8\% contamination rate). These marginal cases, caused by the limitations in reconstructing 2D emission line maps from galactic star-forming regions or the neighboring sources, can be further refined through visual inspection of the reference images and the emission-line maps. Notably, all 11 ELGs from the ZFIRE survey were excluded from the QSO selection region.

\subsection{New QSO candidates in 3D-HST survey} \label{sec:candidates}
In this section, we selected 19 new QSO candidates from the forward modeling results of 1,046 sources with two ``broad'' emission lines and 5,772 sources with one ``broad'' emission line. These new QSO candidates are not included in the SDSS DR16 QSO or Milliquas ``Q'' catalog. The details of these candidate sources are summarized in Table \ref{tab:candidates_infor}. The fifth and sixth rows of Figure \ref{fig: fitting_re} present representative 2D emission-line maps and fitting results for our new QSO candidates, showing H$\alpha$ and \rm{He\textsc{I} 10830\r{A}} features respectively. These candidates demonstrate the same characteristic improvement in modeling extended emission-line wings as observed in both known QSOs (first row) and simulated cases (last row).

For sources with two or more emission lines, the input redshift range for Grizli is centered around the ASERA-detected redshift with a tolerance of 0.15. We than calculate the $\Delta BIC$ and $ratio_Q$ value from the best modeling results, and select seven candidates with \rm{H$\beta$} and \rm{H$\alpha$} emission lines detected in the G141 grism spectra within a redshift range of 1.2-1.6. 


We use the photo-$z$ results from Section \ref{subsec: grizli_intro} to define the redshift range for the modeling of the one-emission-line sources, centering the redshift on the photo-$z$ value with a range of $0.15\times(1+z_{photo})$, and then calculate the $\Delta BIC$ and $ratio_Q$ value from the best modeling results. For all sources that met the criteria were further visually inspected with the F140W image and G141 grism to remove possible ELG features from star-forming knots and nearby sources, resulting in the selection of 12 final candidates. Among these 12 candidates, 11 have the \rm{H$\alpha$} emission line detected in the G141 grism spectra within a redshift range of 0.85–1.38. One candidate exhibits a redshift near 0.12 in both the Grizli fitting and photo-$z$ results, suggesting that the broad \rm{He\textsc{I} 10830\r{A}} line may be detected in the G141 spectrum (see the sixth row in Figure \ref{fig: fitting_re}).

In total, we select 19 candidates spanning AB magnitudes of 20-23 in the F140W band. As shown in Table \ref{tab:candidates_infor}, 7/19 of the candidates have \rm{H$\beta$} and \rm{H$\alpha$} emission lines detected, while 12/19 of the candidates have \rm{H$\alpha$} or \rm{He\textsc{I} 10830\r{A}} line detected in the G141 grism spectra.

\begin{table*}
	\centering
	\caption{19 QSO candidates selected by G141 slitless spectroscopy}
	\label{tab:candidates_infor}
	\begin{tabular}{cccccccccc} 
		\hline
		Name & RA & DEC & z & $\Delta$BIC & $ratio_{Q}$ & F140W mag & $f_{AGN}$ & Emission line & Notes\\
		\hline
		AEGIS-18324 & 214.75613 & 52.7527 & 0.851 & -126.69 & 23.48 & 20.13 & 0.67 & \rm{H$\alpha$} & DESI EDR \& X-ray\\
		AEGIS-38195 & 214.67162 & 52.77345 & 1.487 & -30.99 & 10.42 & 21.74 & 0.86 & \rm{H$\beta$}, \rm{H$\alpha$} & DESI EDR \& X-ray\\
        COSMOS-4305 & 150.18568 & 2.22236 & 1.214 & -4.3 & 6.62 & 21.85 & 0.0 & \rm{H$\beta$}, \rm{H$\alpha$} &  \\
        COSMOS-11337 & 150.11957 & 2.29579 & 1.368 & -56.2 & 42.16 & 21.80 & 0.96 & \rm{H$\alpha$} & X-ray \\
        COSMOS-22536 & 150.06851 & 2.40944 & 1.563 & -154.03 & 17.04 & 21.61 & 0.80 & \rm{H$\beta$}, \rm{H$\alpha$} &  \\
        GOODSN-5487 & 189.02394	& 62.16752 & 0.942 & -12.21	 & 4.71 & 21.03 & 0.54 & \rm{H$\alpha$} & X-ray  \\
        GOODSN-30204 & 189.31947 & 62.29259 & 1.153 & -722.53 & 153.42 & 20.96 & 0.41 & \rm{H$\alpha$} & X-ray \\
        GOODSN-31179 & 189.30063 & 62.29832 & 1.252 & -14.63 & 4.85 & 21.46 & 0.89 & \rm{H$\alpha$} &  X-ray\\
        GOODSN-37623 & 189.27605 & 62.36018 & 0.902 & -121.26 & 12.2 & 20.07 & 0.89 & \rm{H$\alpha$} & DESI EDR \& X-ray\\
        GOODSS-6677 & 53.19818 & -27.87867 & 1.247 & -7.32 & 4.54 & 21.48 & 0.07 & \rm{H$\beta$}, \rm{H$\alpha$} &  \\
        GOODSS-12262 & 53.18836 & -27.85353 & 1.257 & -3.2 & 4.58 & 22.50 & 0.05 & \rm{H$\alpha$} &  \\
        GOODSS-13301 & 53.05835 & -27.85019 & 0.114 & -40.3 & 5.93 & 20.80 & 0.44 & \rm{He\textsc{I} 10830\r{A}} & X-ray \\
        GOODSS-20651 & 53.14988 & -27.814 & 1.310 & -58.73 & 8.04 & 21.19 & 0.11 & \rm{H$\beta$}, \rm{H$\alpha$} & AGN in \cite{2022ApJ...941..191L} \& X-ray\\
        GOODSS-43266 & 53.14284 & -27.70694 & 1.091 & -13.51 & 4.9 & 20.31 & 0.0 & \rm{H$\alpha$} &  \\
        UDS-13108 & 34.47922 & -5.23364 & 1.309 & -51.85 & 8.07 & 20.55 & 0.95 & \rm{H$\beta$}, \rm{H$\alpha$} & X-ray\\
        UDS-14583 & 34.36742 & -5.2285 & 1.192 & -33.75 & 6.68 & 20.75 & 0.04 & \rm{H$\alpha$} &  \\
        UDS-21513 & 34.38555 & -5.20501 & 1.382 & -54.39 & 7.86 & 21.06 & 0.36 & \rm{H$\alpha$} & X-ray\\
        UDS-26875 & 34.41814 & -5.18799 & 1.365 & -20.39 & 5.16 & 21.02 & 0.93 & \rm{H$\beta$}, \rm{H$\alpha$} & X-ray\\
        UDS-34473 & 34.57496 & -5.16168 & 1.194 & -26.56 & 5.75 & 20.95 & 0.33 & \rm{H$\alpha$} &  \\
	   \hline
	\end{tabular}
\end{table*}

\subsection{Validation of QSO Candidates}
\subsubsection{Candidates confirmed by other projects}
\begin{figure}
\includegraphics[width=\columnwidth]{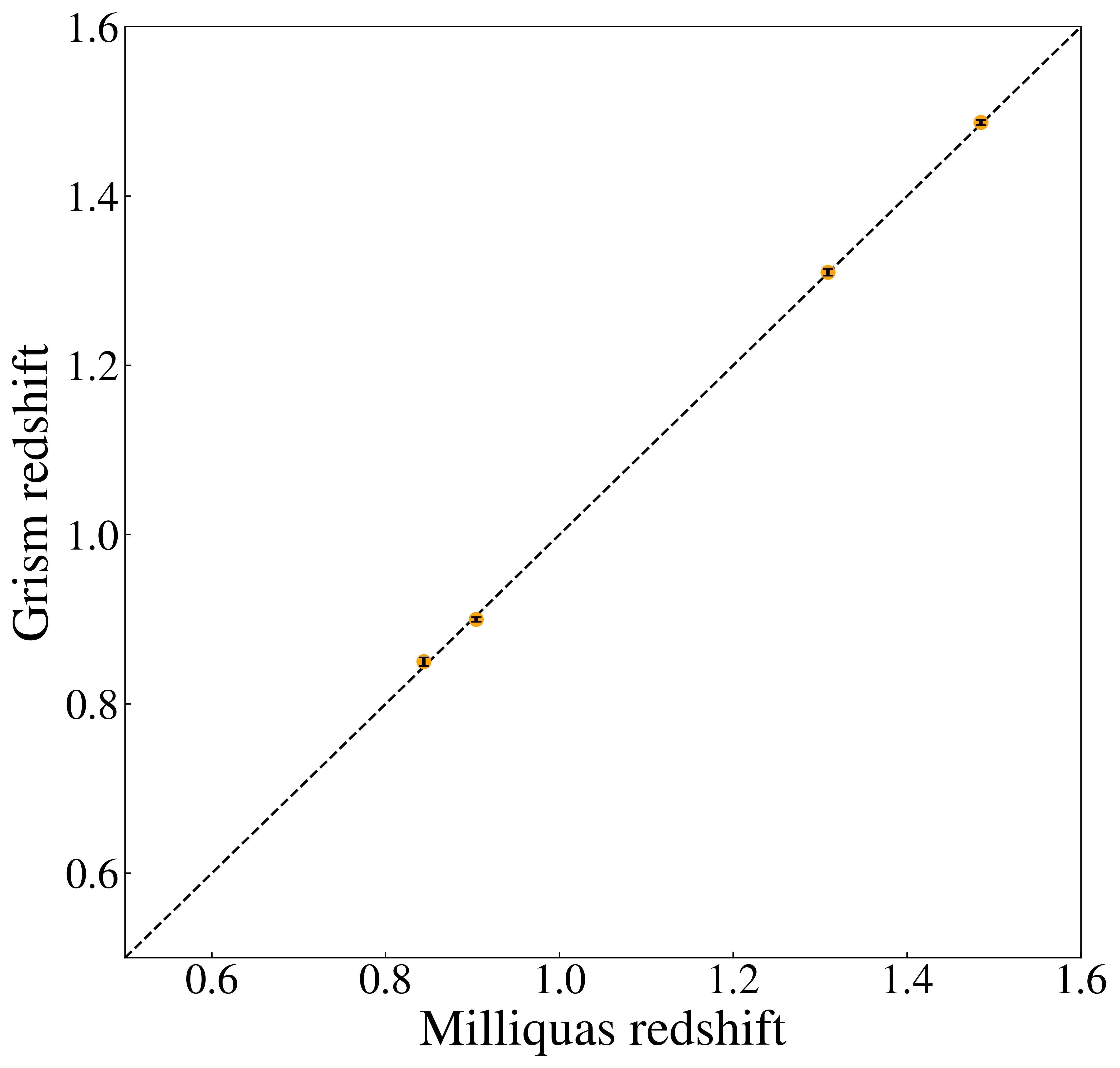}
\caption{Redshift comparison between the redshifts from Grizli modeling and those from other projects for our candidates. For reference, the black dashed line represents an equal redshift.
\label{fig: compare_new_dis}}
\end{figure}

When we started our project, we used the v7.5 Milliquas catalog. Several new QSOs in these deep fields have been discovered very recently by other teams throughout this two-year project period. The DESI early data release confirmed three candidates: AEGIS-18324, AEGIS-38195, and GOODSN-37623 \citep{2024AJ....168...58D}. Additionally, \cite{2022ApJ...941..191L} identified AGNs in the GOODSS field through broad-band SED fitting, including our candidate GOODSS-20651. Figure \ref{fig: compare_new_dis} shows the redshift comparison for these four sources, and the redshifts obtained from the grism data closely matching those from other projects.

\subsubsection{Chandra X-ray detection}
We cross-match our candidates with the Chandra v2.1 catalog \citep{2024arXiv240710799E}, finding that 12 out of 19 of our candidates have Chandra X-ray detections.

For these 12 sources with X-ray observations, their high X-ray luminosities serve as crucial evidence supporting their AGN nature. More generally, the X-ray characters and the broad emission line features in slitless spectroscopy exhibit different preferences in AGN identification. X-ray-selected AGN samples encompass both classical Type I AGN and optically obscured Type II AGN \citep{2011A&A...534A.110L,2016A&A...594A..72P,2021A&A...653A..70M}, where the latter do not display broad Balmer emission line wings and thus lack significant broad-line features in slitless spectra. On the other hand, AGNs with X-ray column densities exceeding $\rm{N_{H}}>10^{24}\rm{cm^{-2}}$ become Compton-thick and are challenging to detect in soft X-ray surveys with insufficient depth \citep{2014MNRAS.443.1999B, 2019A&A...629A.133C}. This may explain why some of our broad-line selected candidates lack actual X-ray flux detections.

\subsubsection{CIGALE SED fitting} \label{subsubsec:cigale}
We employed CIGALE\footnote{\url{https://cigale.lam.fr/2022/07/04/version-2022-1/}} \citep[Code Investigating GALaxy Emission;][]{2019A&A...622A.103B} to perform the detailed SED modeling for our candidates with considering the energy balance between the UV/optical and infrared. The 2022 version of CIGALE \citep{2022ApJ...927..192Y} is able to handle various parameters such as star formation history (SFH), single stellar population (SSP), attenuation laws, AGN emission, dust emission, and X-ray emission. This flexibility has made CIGALE adopted in numerous AGN studies \citep[i.e.,][]{2018ApJ...866..140Y,2019ApJS..243...15T}.

\begin{figure}
\includegraphics[width=\columnwidth]{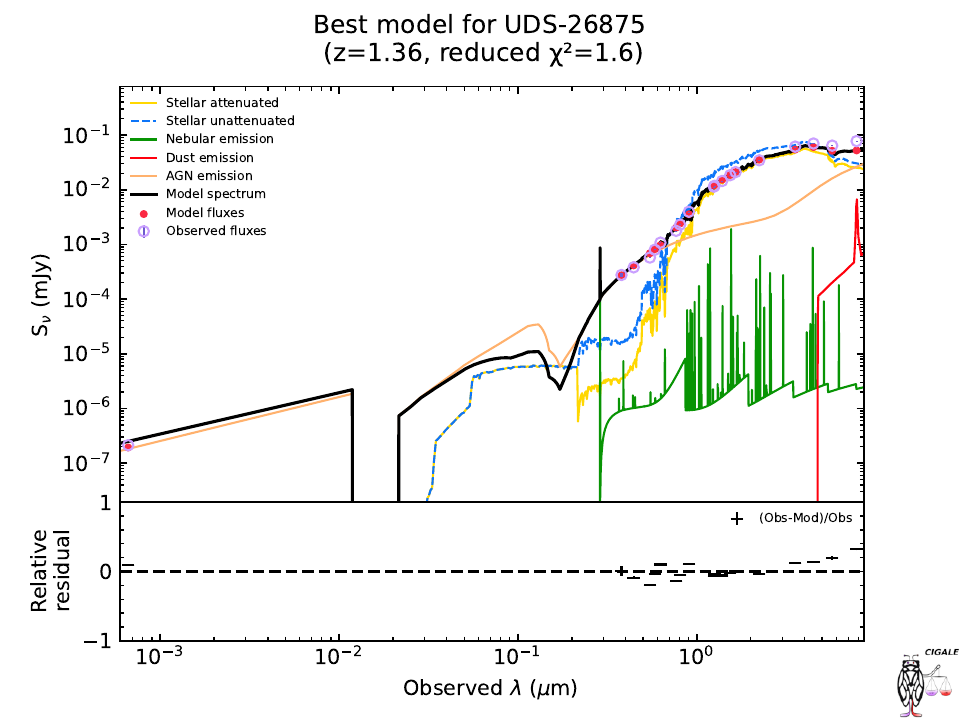}
\caption{CIGALE fitting results for one of our candidates, UDS-26875. The orange line represents the AGN component of this source.
\label{fig: cigale_results}}
\end{figure}

The input for CIGALE includes multi-band photometry from UV to mid-infrared provided by 3D-HST, along with X-ray data for those sources that have detection in Chandra. We adopt the same template list as \citet{2019ApJS..243...15T} for the fitting process, with a systematic error of 0.1 magnitude added during the fitting process. Galactic extinction corrections were applied to all photometry bands using the dust map from \cite{2014A&A...571A..11P}. The ``AGN fraction'' column in Table \ref{tab:candidates_infor} shows the F140W band AGN fraction obtained from the CIGALE fitting results. Almost all candidates (17/19) display an AGN component in the multi-band SED fitting. Figure \ref{fig: cigale_results} illustrates an example of the CIGALE fitting results.

CIGALE assumes that the X-ray detections are predominantly attributable to an AGN component, a reasonable assumption for sources with high X-ray fluxes. It should be noted, however, that AGN identification becomes suggestive but not certain at lower $L_X$. For instance, the faintest AGN identified through X-ray selection in \citet{2019A&A...622A..29M} has luminosities around $10^{42.5}\rm{erg~s^{-1}}$, while one of the AGN definitions adopted by \citet{2022MNRAS.510.4556B} requires X-ray luminosity $10^{42} \rm{erg~s^{-1}}$. From the Chandra X-ray flux catalog, three sources (GOODSS-13301, GOODSS-20651, and UDS-21513) have lower-limit $L_X$ values below $10^{42}\rm{erg~s^{-1}}$. We therefore performed additional CIGALE fittings excluding X-ray data for these sources, resulting in AGN fraction changes from 0.44 to 0.56, 0.11 to 0.08, and 0.42 to 0.36, respectively, which represent minor changes in the AGN fraction results.

Although most of our sources exhibit an AGN fraction in the CIGALE fitting, we emphasize that the derived AGN fractions may have significant systematic uncertainties. Firstly, the multi-band fluxes used in SED fitting were retrieved from \citet{2016ApJS..225...27M} results re-calibrated with previous catalogs spanning about 15 years of observations, which could be affected by AGN variability. Additionally, the ground-based telescopes have lower image resolutions compared to HST observation, leading to wavelength-dependent aperture photometry in the \citet{2016ApJS..225...27M} catalog. This mismatch may introduce photometric inconsistencies. Finally, degeneracies between the host galaxy and AGN templates can arise due to limited wavelength coverage in our photometric data. Mid-infrared (MIR) data provide critical constraints on torus properties \citep{2015ARA&A..53..365N} and help disentangle degeneracies between AGN and stellar components in our fittings.

\section{Properties of the New QSO candidates in the 3D-HST survey} \label{sec: can_properties}
\subsection{Color comparison with QSOs in SDSS \& Milliquas} \label{subsec:compare}
\begin{figure}
\includegraphics[width=\columnwidth]{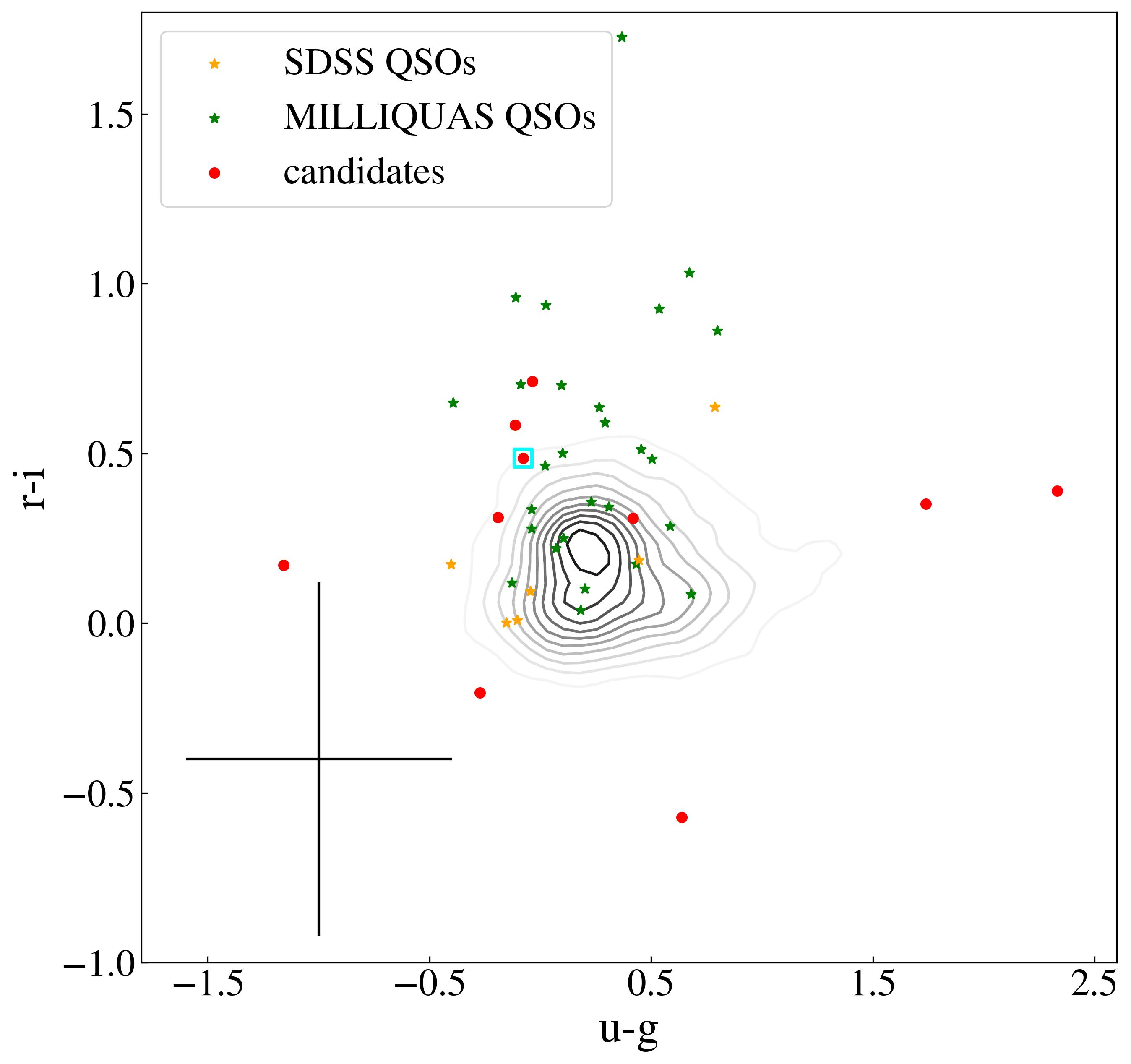}
\caption{The rest-frame UV color comparison between our candidates (red dots), SDSS QSOs (orange stars), and Milliquas QSOs (green stars) in the 3D-HST field. The black contour shows all the SDSS QSOs distribution given by SDSS photometry. The cyan square shows the source, which is further confirmed by DESI EDR. The black cross on the left lower corner shows the typical color error of 3D-HST sources.
\label{fig: compare_color}}
\end{figure}

In this study, our selection method primarily relies on the emission line properties of slitless spectra in the G141 band. This method differs from the conventional QSO selection methods based on colors. Therefore, it is necessary to compare the color properties of our sources with those of previous QSO samples. Figure \ref{fig: compare_color} compares u-g versus r-i colors of our new candidates with previous known QSOs in the 3D-HST field from the CFHT/MegaCam, VLT/VIMOS, and Subaru/Suprime-Cam photometry \citep[see more details in][]{2014ApJS..214...24S}. Despite the 3D-HST photometry catalog having larger color uncertainties, our sample exhibits a broader range of colors compared to the SDSS QSO sample.

\begin{figure}
\includegraphics[width=\columnwidth]{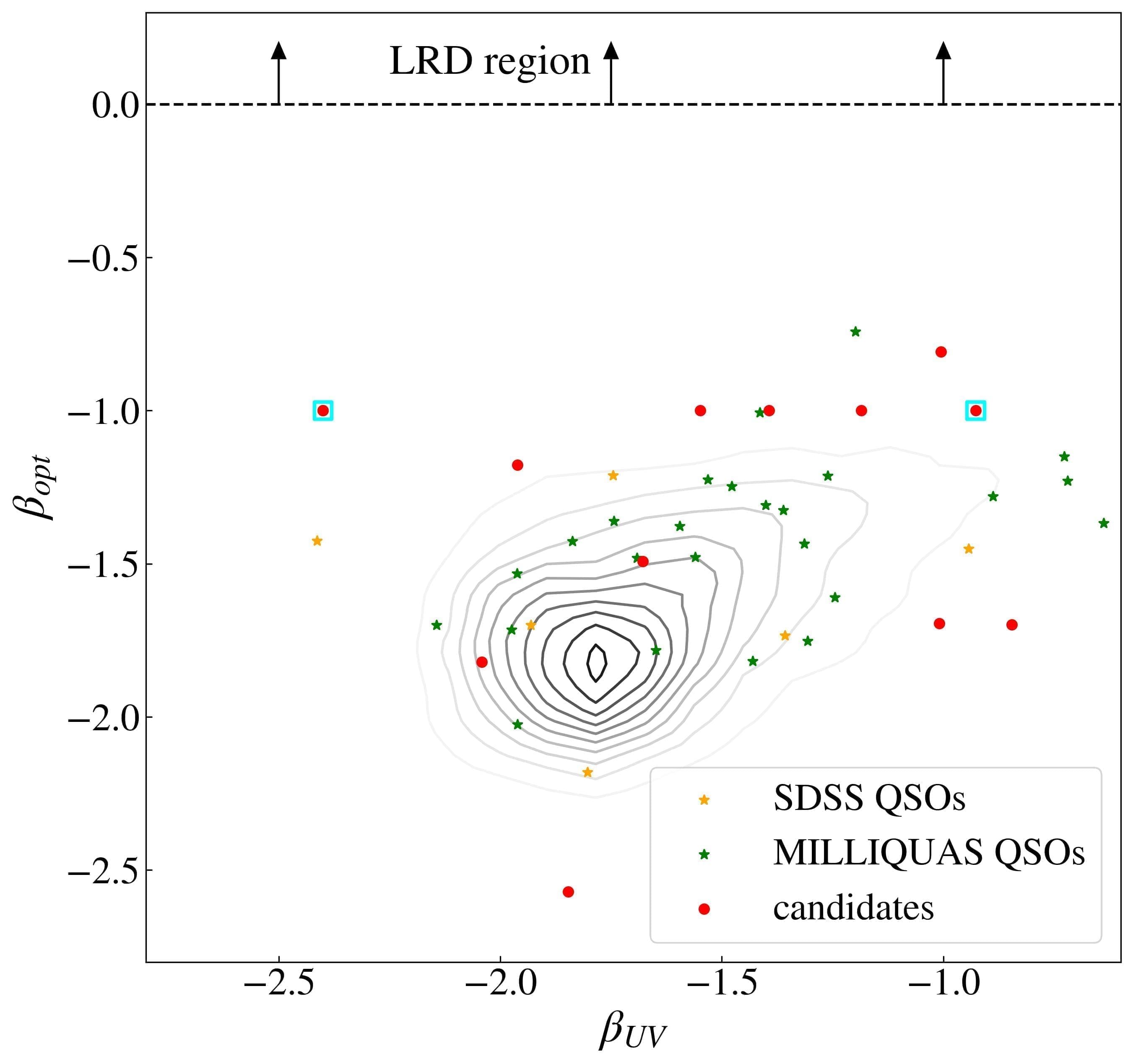}
\caption{The rest-frame UV and optical slope comparison between our candidates (red dots), SDSS QSOs (orange stars), and Milliquas QSOs (green stars) in the 3D-HST field. The black contour shows all the SDSS QSOs distribution calculated by SDSS and 2MASS photometry. The cyan square shows the source, which is further confirmed by DESI EDR. The dotted dash line represented the little red dots (LRDs) region from \citep{2024arXiv240403576K}.
\label{fig: compare_slope}}
\end{figure}

Recently, JWST reveals an abundant population of red compact sources at $z>4$, nicknamed as `little red dots' \citep[LRDs,][]{2024ApJ...963..128B,2023arXiv230607320L,2024ApJ...963..129M}. Their number density is 10-100 times higher than the extrapolation from quasars luminosity functions \citep{2023ApJ...954L...4K}. LRDs typically exhibit V-shape SED with a redder optical continuum compared to traditional QSOs \citep{2024ApJ...964...39G,2023arXiv231203065K}. Most of the LRDs also detected broad H$\alpha$ lines in the JWST spectroscopy data \citep{2023ApJ...959...39H,2024arXiv240403576K}. Figure \ref{fig: compare_slope} compares the UV and optical slopes of our new candidates with previously known QSOs and the LRDs region. The slope definitions and fitting procedures follow the description in \cite{2024arXiv240403576K}. Considering our sample's redshift distribution is mainly around 0.8-1.6, we performed the UV slope fitting using the g, r, and i band photometry from the CFHT/MegaCam and Keck/LRIS. The optical slope fitting uses the z band photometry from the CFHT/MegaCam and Subaru/Suprime-Cam; J and H bands photometry from CFHT/WIRCam, Subaru/MOIRCS, VLT/ISAAC, and UKIRT/WFCAM \citep[see more details in][]{2014ApJS..214...24S}. The UV slope distribution of our candidates aligns closely with the SDSS sample, while the optical slopes are redder than the SDSS QSOs. However, our candidates' color and slope distributions fall within the range of the Milliquas QSO sample and do not overlap with the LRD selection criteria.

\subsection{Physical Properties} \label{subsec: phy-properties}
To systematically analyze the physical properties of our newly identified QSO candidates, we utilize QSOFITMORE \citep[version 1.1.0;][]{yuming_fu_2021_5810042} - a customized Python package built upon the PyQSOFit framework \citep[][]{2018ascl.soft09008G}. The G141 1D spectra obtained through Grizli optimal extraction were first corrected for Galactic extinction using the \cite{2014A&A...571A..11P} dust maps and the \cite{2019ApJ...877..116W} extinction law. For low-redshift sources (z < 1.16), we implement principal component analysis (PCA) decomposition through PyQSOFit to disentangle the host galaxy and QSO components \citep{2004AJ....128..585Y}.

The spectral analysis proceeds through two key stages:
\begin{itemize}
    \item Continuum modeling using a third-order polynomial combined with Fe~\textsc{II} templates ($f_{Fe\textsc{II}}$) to establish the pseudo-continuum ($f_{cont}$), with emission line regions masked during this process.
    \item Multi-Gaussian fitting of emission line profiles after continuum subtraction, adopting the parameterization scheme developed by \cite{2025ApJ...980..223P} for CSST slitless spectroscopy simulations.
\end{itemize}

\begin{figure*}
    \centering
    \includegraphics[width=0.90\textwidth]{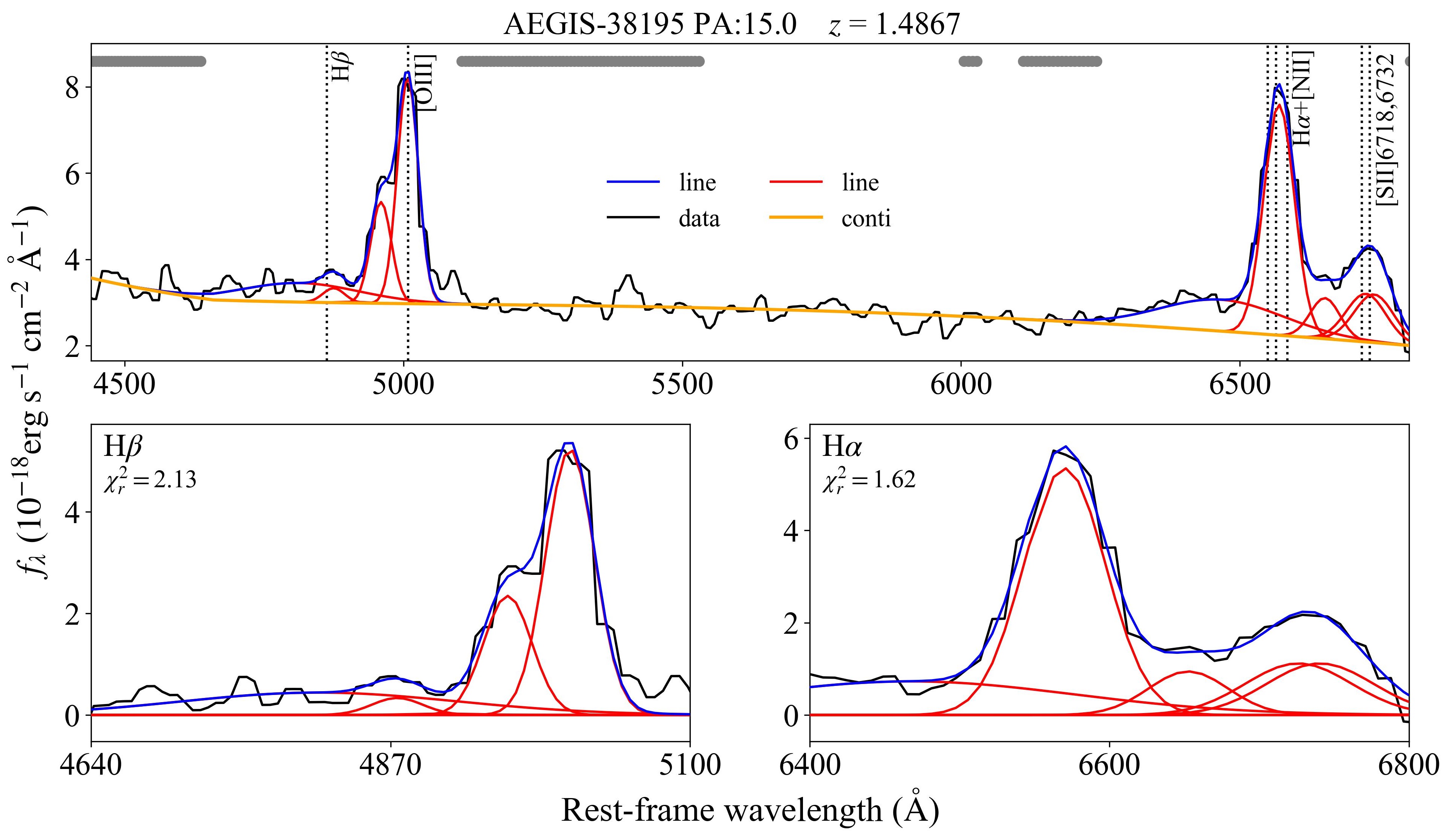}
    \caption{QSOFITMORE fitting results of a G141 1D spectum. The black lines denote the total extracted spectrum, the yellow lines denote the continuum, the blue lines denote the total flux of different emission lines, and the red lines denote each Gaussian component.}
    \label{fig: fitting-example}
\end{figure*}

Uncertainty quantification is performed through Monte Carlo simulations. Figure \ref{fig: fitting-example} shows a typical fitting example of the G141 spectra. Detailed methodology follows the standardized procedures outlined in \cite{2011ApJS..194...45S}, \cite{2019ApJS..241...34S}, and \cite{2022ApJS..261...32F}.

To ensure sufficient data points for multi-Gaussian fitting for emission lines, we implement flux-conserving spectral resampling using the \verb|FluxConservingResampler| function in the \verb|specutils| package. This process achieves a wavelength sampling density equivalent to a resolution of 800. Furthermore, to account for slitless spectroscopy's line-broadening effects, we extend the upper limit of the fitting for narrow line widths to the same limit of the corresponding broad lines.

The initial QSOFITMORE outputs exhibit systematically overestimated FWHM measurements due to point spread function (PSF) convolution effects. For the WFC3 G141 grism, the dispersion is 46.5 \r{A}~$\rm pixel^{-1}$ ($R\sim130$) in the primary (+1 st) spectral order, with the PSF pattern having an FWHM between 1.02-1.22~$\rm pixel$ \citep[WFC3 Handbook;][]{2010wfci.book....3D}. We implement spectral resolution correction through:
\begin{equation}
{\rm FWHM_{int}} = \sqrt{{\rm FWHM_{obs}^2 - FWHM_{psf}^2}},
\end{equation}
where $\rm FWHM_{psf} \sim 1000km/s$ corresponds to HST's instrumental broadening. As tabulated in Table 2 (Column "FWHM"), the corrected broad H$\alpha$/He~\textsc{I} line widths range 2000-7000 km/s. These values align with AGN broad-line characteristics \citep[FWHM > 1200 km/s; ][]{2005AJ....129.1783H,2022ApJS..261...24L,2022ApJS..261....7D}.

With the continuum luminosity serving as a proxy for the broad-line region size \citep[i.e., the R-L relation; e.g.,][]{2004A&A...424..793W, 2006ApJ...644..133B, 2016ApJ...825..126D} and the broad-line width as a proxy for the virial velocity, we estimate the single epoch virial black hole masses ($\rm M_{BH}$
) using the empirical mass-scaling relations calibrated with the reverberation mapping (RM) masses \citep[e.g.,][]{2004MNRAS.352.1390M, 2006ApJ...641..689V}. We adopt the H$\alpha$-based estimator from \citet{2005ApJ...630..122G} and the He~\textsc{I}-based estimator from \citet{2013MNRAS.432..113L}:
\begin{equation}
\begin{split}
    \mathrm{l}&\mathrm{og}(M_{\mathrm{BH}}/M_{\odot})  \\
    &=\mathrm{log} \left[\left(\frac{\mathrm{FWHM}(\mathrm{H}\alpha)}{\mathrm{km~s^{-1}}}\right)^{2.06}\left(\frac{L_{\mathrm{H}\alpha}}{10^{42}\mathrm{erg~s^{-1}}}\right)^{0.55}\right]+0.12,
\end{split}
\end{equation}
\begin{equation}
\begin{split}
    \mathrm{l}&\mathrm{og}(M_{\mathrm{BH}}/M_{\odot})  \\
    &=\mathrm{log} \left[\left(\frac{\mathrm{FWHM}(\mathrm{NIR})}{\mathrm{km~s^{-1}}}\right)^{1.76}\left(\frac{L_{1\mu m}}{10^{45}\mathrm{erg~s^{-1}}}\right)^{0.44}\right]+2.41,
\end{split}
\end{equation}
Where the total H$\alpha$ line luminosity and 1$\mu m$ monochromatic luminosity are derived from spectral fitting. The uncertainties in $\rm M_{BH}$ are propagated from MC errors of the line widths and the monochromatic continuum luminosity. We compute the F160W band luminosity, convert it to the rest-frame wavelength, and apply the bolometric corrections from \cite{2006ApJS..166..470R} to obtain the bolometric luminosity ($L_{bol}$). The Eddington ratio ($\lambda_{\rm{Edd}}$) is then calculated as:
\begin{equation}
\lambda_{\rm Edd} = \frac{L_{\rm bol}}{L_{\rm Edd}} = \frac{L_{\rm bol}}{1.26\times10^{38},{\rm erg,s^{-1}}} \frac{M_{\odot}}{M_{\rm BH}}.
\end{equation}

We summarized the estimated physical properties above in Table \ref{tab: candidates_phy_properties}. It is noteworthy that the dispersion of $\rm M_{BH}$ and ($\lambda_{\rm{Edd}}$) estimations does not account for the 0.2-0.3 dex systematic uncertainty inherent to the single-epoch virial method and the bolometric corrections. Our new sample exhibits black hole masses of $10^{6.9}-10^{8.3} M_{\odot}$, systematically lower than SDSS quasars, with Eddington ratios approximately 0.5 dex lower.

\begin{table*}
	\centering
	\caption{Physical properties of the 19 QSO candidates}
	\label{tab: candidates_phy_properties}
	\begin{tabular}{cccccc} 
		\hline
		Name & $\rm L_{5100 \text{\r{A}}}$ & $\rm log(L_{H\alpha})$ & FWHM & $\rm M_{BH}$ & $\lambda_{\rm{Edd}}$ \\
             & (log[erg/s]) & (log[erg/s]) & (km/s) & ($\rm log(M_{\odot})$) & ($\rm log(\lambda_{Edd})$) \\
		\hline
		AEGIS-18324 &  & 42.54 & 7114$\pm$197 & 8.35$\pm$0.03 & -2.18$\pm$0.04 \\
            AEGIS-38195 & 43.32 & 42.98 & 3061$\pm$167 & 7.84$\pm$0.05 & -1.55$\pm$0.06 \\
            COSMOS-4305 & 42.99 & 42.39 & 4053$\pm$183 & 7.77$\pm$0.04 & -1.86$\pm$0.05 \\
            COSMOS-11337 & 42.74 & 42.76 & 3311$\pm$386 & 7.79$\pm$0.10 & -1.69$\pm$0.11 \\
            COSMOS-22536 & 43.41 & 43.15 & 3349$\pm$2143 & 8.01$\pm$0.58 & -1.74$\pm$0.58 \\
            GOODSN-5487 &  & 42.28 & 3458$\pm$159 & 7.57$\pm$0.05 & -1.64$\pm$0.05 \\
            GOODSN-30204 &  & 43.05 & 4033$\pm$373 & 8.12$\pm$0.08 & -1.98$\pm$0.09 \\
            GOODSN-31179 & 43.07 & 42.53 & 4240$\pm$203 & 7.89$\pm$0.05 & -1.79$\pm$0.05 \\
            GOODSN-37623 &  & 42.73 & 4134$\pm$200 & 7.97$\pm$0.05 & -1.78$\pm$0.05 \\
            GOODSS-6677 & 43.27 & 42.78 & 2818$\pm$34 & 7.66$\pm$0.01 & -1.58$\pm$0.03 \\
            GOODSS-12262 & 42.85 & 42.15 & 4460$\pm$401 & 7.72$\pm$0.08 & -2.03$\pm$0.09 \\
            GOODSS-13301 & 41.15$^1$ & 39.87$^2$ & 3362$\pm$1750 & 6.92$\pm$0.44 & -3.01$\pm$0.44 \\
            GOODSS-20651 & 43.44 & 42.73 & 3430$\pm$233 & 7.81$\pm$0.07 & -1.57$\pm$0.07 \\
            GOODSS-43266 &  & 42.36 & 4985$\pm$317 & 7.93$\pm$0.06 & -1.54$\pm$0.07 \\
            UDS-13108 & 43.54 & 42.91 & 3728$\pm$109 & 7.98$\pm$0.03 & -1.49$\pm$0.04 \\
            UDS-14583 & 43.38 & 42.51 & 5165$\pm$254 & 8.05$\pm$0.04 & -1.73$\pm$0.05 \\
            UDS-21513 & 43.39 & 42.70 & 4766$\pm$387 & 8.08$\pm$0.08 & -1.69$\pm$0.08 \\
            UDS-26875 & 43.45 & 42.63 & 3859$\pm$133 & 7.86$\pm$0.03 & -1.45$\pm$0.04 \\
            UDS-34473 & 43.35 & 42.53 & 4326$\pm$228 & 7.90$\pm$0.05 & -1.66$\pm$0.06 \\
	   \hline
	\end{tabular}\\
        \footnotesize{$^1$ This monochromatic luminosity is measured at 1$\mu m$}\\
        \footnotesize{$^2$ This emission line luminosity is measured by \rm{He\textsc{I} 10830\r{A}}}\\
\end{table*}

\subsection{Flux ratio of emission line} \label{subsec: flux_ratio}
For sources with simultaneous detections of H$\alpha$, H$\beta$ and [O~\textsc{III}] in the G141 spectra, we calculated their flux ratios, as listed in Table \ref{tab: candidaes_line_ratio}.

The H$\alpha$ to H$\beta$ flux ratios of our new QSO sample range from 2.3 to 4.5, consistent with previous studies. For instance, the SDSS DR7 results from \citet{2011ApJS..194...45S} show a similar range of 2.3–4.4, and \citet{2017MNRAS.467..226G} derived a median value of 2.72 after applying reddening corrections. These observed ratios are also broadly consistent with the theoretical predictions from the CASE B model \citep{1938ApJ....88...52B}.

Due to the larger uncertainties in the H$\beta$ and [O~\textsc{III}] flux measurements, we calculated the H$\alpha$-to-[O~\textsc{III}] flux ratios instead of H$\beta$-and-[O~\textsc{III}] flux ratios. Since the [O~\textsc{III}] 4959, 5007\r{A} doublet is unresolved in the G141 1D spectra, we summed the fluxes of both lines to calculate the total [O~\textsc{iii}] flux. Our sample exhibits a wide range of 1.2–11 for this ratio, reflecting the intrinsic diversity of [O~\textsc{iii}] emission strength in AGNs—a phenomenon linked to "Eigenvector 1" \citep{2014Natur.513..210S, 2022MNRAS.514.1595W}. For comparison, the SDSS DR7 catalog shows H$\alpha$-to-the [O~\textsc{III}] flux ratios spanning 1.5–14, consistent with our results. It is worth noting that the [O~\textsc{III}] line calculated here may contain contributions from the host galaxy (see Section \ref{sec:discuss} and Figure \ref{fig: Ha_OIII_compare}), therefore the H$\alpha$-to-[O~\textsc{III}] ratio should practically be considered as a lower limit.

\begin{table}
	\centering
	\caption{Emission line ratios of new QSO candidates}
	\label{tab: candidaes_line_ratio}
	\begin{tabular}{ccc} 
	    \hline
	    Name & $\rm Ratio_{H\alpha, H\beta}$ & $\rm Ratio_{H\alpha, [O III]}$\\
	    \hline
            AEGIS-38195 & $4.10\pm1.01$ & $2.11\pm0.19$ \\
            COSMOS-4305 &  & $1.93\pm0.54$ \\
            COSMOS-22536 & $4.10\pm6.22$ & $5.34\pm8.10$ \\
            GOODSN-31179 &  & $11.55\pm3.04$ \\
            GOODSS-12262 &  & $2.82\pm0.79$ \\
            GOODSS-20651 & $8.96\pm3.78$ & $1.83\pm0.32$ \\
            GOODSS-6677 & $4.55\pm4.75$ & $1.10\pm0.12$ \\
            UDS-13108 & $3.62\pm9.29$ & $5.92\pm1.04$ \\
            UDS-26875 & $2.39\pm0.30$ & $1.94\pm0.15$ \\
            SDSS QSOs$^*$ & 2.3-4.4 & 1.5-14 \\
	    \hline
	\end{tabular}\\
        \footnotesize{$^*$ From SDSS DR7 catalog by \citet{2011ApJS..194...45S}}\\
\end{table}

\subsection{Luminosity function}
Most of our newly discovered candidates have H$\alpha$ emission lines within the G141 band, corresponding to z = 0.8-1.6. In this section, we briefly discuss their impact on the bolometric LF.

For calculating the LF in the 3D-HST fields, we constructed a quasar sample within the redshift range 0.8–1.6, incorporating QSOs confirmed by the SDSS survey, DESI early data release, follow-up observations of X-ray detections, and our newly identified candidates. This combined sample consists of 31 QSOs, including 16 previously known QSOs and 18 new candidates (3 of which overlap with DESI QSO). These quasars were selected using different methods with varying completeness levels. For example, SDSS and DESI primarily use color selection, which introduces a bias toward certain color ranges; X-ray follow-up observations may miss quasars with weak X-ray fluxes; and our sample excludes sources affected by contamination. As a result, we did not apply the completeness corrections, and the sample size should be considered as a lower limit for the total quasar population. Due to the uncertainties in the host galaxy contributions, we exclude AGNs in the Milliquas catalog, such as radio-loud AGNs from VLA and QSOs identified through HST variability. We use the method in Section \ref{subsec: phy-properties} to obtain $L_{bol}$ of these QSOs. We then divide all QSOs into two luminosity ranges: 10$^{44}$ - 10$^{44.5}$ erg/s and 10$^{44.5}$ - 10$^{45}$ erg/s.

\begin{figure}
\includegraphics[width=\columnwidth]{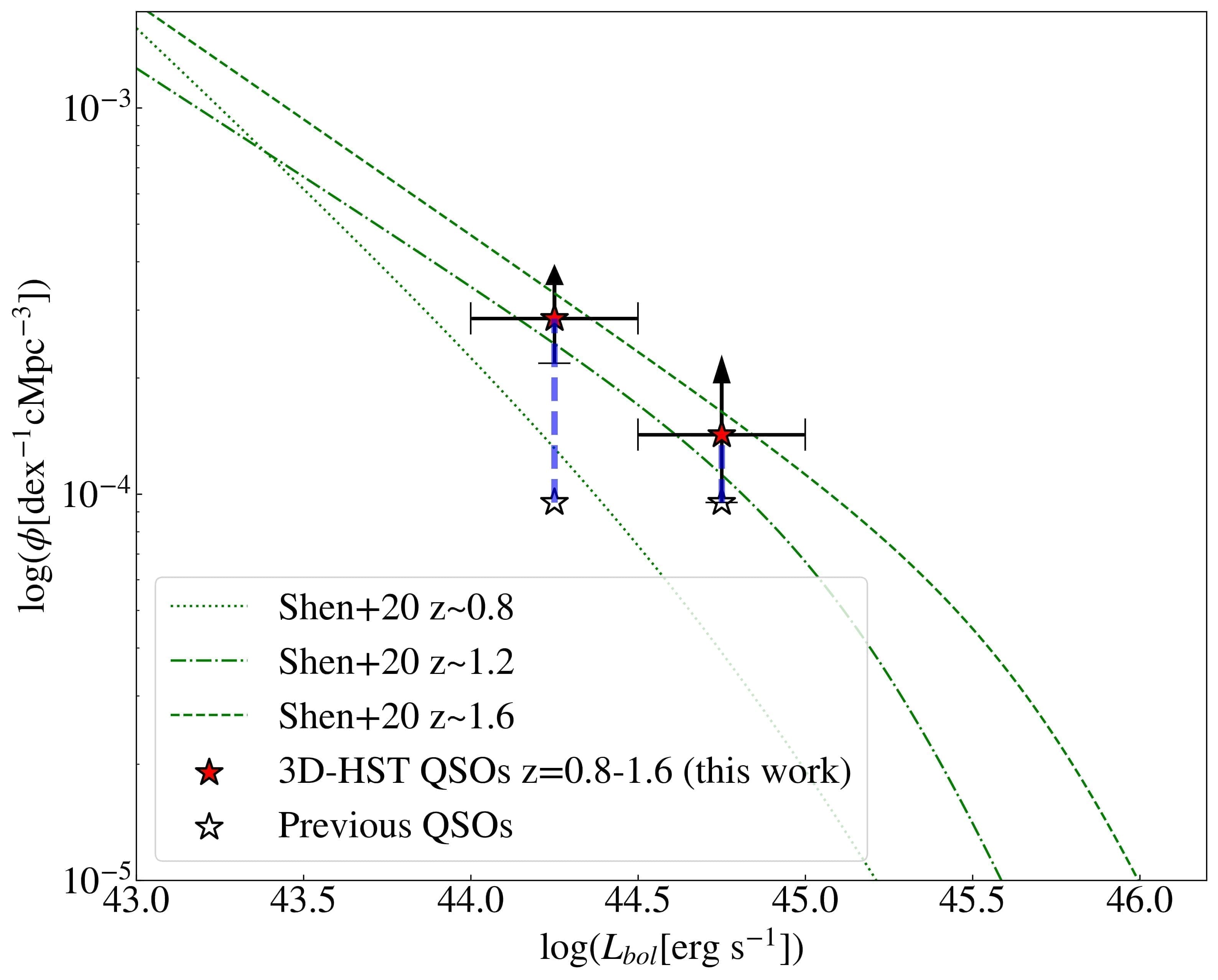}
\caption{The bolometric luminosity function (LF) of QSOs from 3D-HST in the redshift range 0.8-1.6. The bolometric luminosities of these samples are estimated from the F160W flux. The green dotted, dash-dotted, and dashed lines represent the bolometric LFs of quasars at z$\sim$0.8, 1.2, and 1.5 from \citet{2020MNRAS.495.3252S}. The black stars with/without red filling represent the sample including/without our QSO candidates.}
\label{fig: LF_compare}
\end{figure}

Figure \ref{fig: LF_compare} shows the comparison of our sample's luminosity function with that of \cite{2020MNRAS.495.3252S} for similar redshifts. Our results are consistent with the extrapolated bolometric luminosity function of quasars at z $\sim$ 1.2-1.6. In the redshift range of 0.8–1.2, the luminosity function exhibits significant variations in shape and our limited sample size is insufficient to fully characterize this evolution. This luminosity function calculation indicates that the new slitless spectra QSO candidates significantly enhance the completeness of QSOs in the 10$^{44}$ - 10$^{44.5}$ erg/s range at redshifts of 0.8-1.6.

\section{Discussion and Future Perspective} \label{sec:discuss}
Given the significant difference between QSOs and ELGs, the criteria in Equations \ref{formula:criteria} and \ref{formula:criteria2} may still allow for a small percentage of ELG contamination. Thus, addressing the balance between completeness and precision is essential when determining the criteria. Table \ref{tab:criteria-trade-off} offers insight into this trade-off, showing that stricter BIC criteria reduce the ELG contamination represented by the DEEP2 surveys but also result in the loss of more QSOs. Under the strictest criteria, nearly all ELGs would be excluded automatically, but almost all QSOs with host galaxy influence will also be eliminated.

In addition to the H$\alpha$ spectral features discussed in Section \ref{sec:separation}, the two-dimensional characteristics of [O \textsc{III}] emission lines in the slitless spectra hold other astrophysical implications. For star-forming galaxies, the [O \textsc{III}] lines enable estimations of gas-phase metallicity and physical properties \citep{2017ApJ...837...89W,2019ApJ...882...94W,2025ApJ...981...96C}. For active galactic nuclei (AGN), [O \textsc{III}] emission provides insights into the narrow-line region physics \citep{2005MNRAS.358.1043B}, while in some AGN it serves as a tracer of gas outflows, allowing quantification of feedback effects on host galaxies \citep{2015A&A...580A.102C,2017A&A...598A.122B,2023ApJ...951L...5Y}. The comparative analysis of H$\alpha$ and [O \textsc{III}] 5007\r{A} 2D emission line maps can help distinguish nuclear activity, as individual emission lines not originating from the AGN's broad-line region would lack this spatial correspondence.

\begin{figure}
\includegraphics[width=\columnwidth]{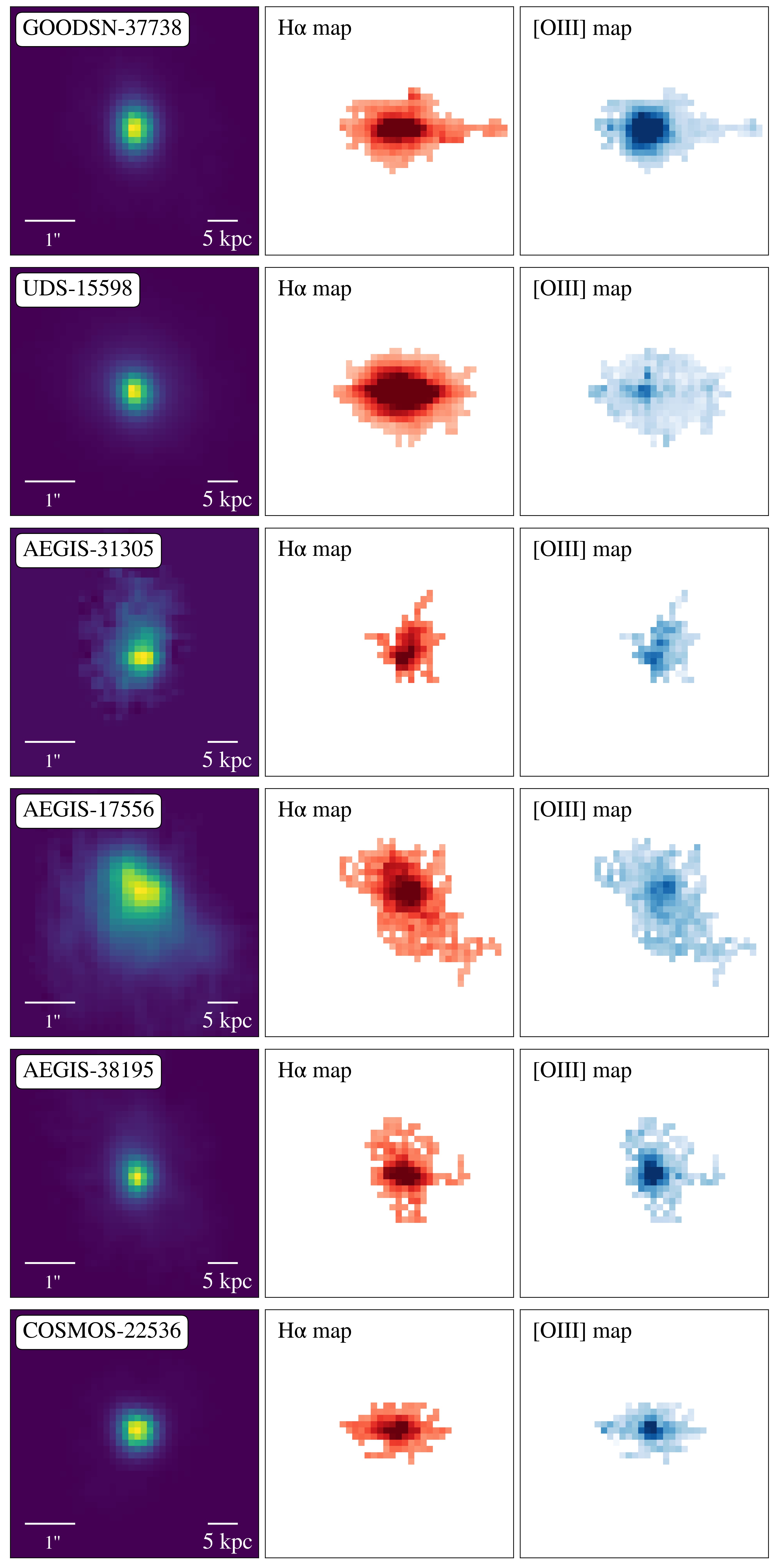}
\caption{Comparison of H$\alpha$ and [O \textsc{III}] 5007\r{A} 2D emission line maps between QSOs and ELGs observed in G141 slitless spectroscopic data. Columns (left to right): F140W reference image with source name annotated in the upper right corner (angular and physical scales shown below); corresponding 2D emission line maps for H$\alpha$ and [O \textsc{III}]. Source classifications from top to bottom: Previously known QSOs (Rows 1-2), spectroscopically confirmed ELGs (Rows 3-4), and newly identified QSO candidates (Rows 5-6).}
\label{fig: Ha_OIII_compare}
\end{figure}

We systematically compared H$\alpha$ and [O \textsc{III}] 5007\r{A} emission lines for known QSOs, emission-line galaxies (ELGs), and new candidates exhibiting more than 20 spatial pixels with S/N larger than 3 in both lines, as shown in Figure \ref{fig: Ha_OIII_compare}. Among the confirmed QSOs, the [O \textsc{III}] 5007\r{A} emission maps reveal two distinct morphologies: extended emission aligned with host galaxy structure when originating from star-forming regions (Row 1: GOODSN-37738), versus compact point-like emission when dominated by the NLR (Row 2: UDS-15598). For ELGs, both H$\alpha$ and [O \textsc{III}] 5007\r{A} emissions display spatial extension, with morphological consistency (Row 3: AEGIS-31305) or divergence (Row 4: AEGIS-17556) depending on localized metallicity variations.

Notably, QSOs exhibit more elongated H$\alpha$ emission along the dispersion direction compared to ELGs, a characteristic preserved in our new QSO candidates (Rows 5-6). Furthermore, no significant [O \textsc{III}] outflow signatures were detected in either confirmed or candidate QSOs, consistent with the low accretion rates observed in this sample \citep{2013MNRAS.433..622M,2016ApJ...817..108W}.

Enhancing the QSO selection efficiency demands multi-PA observations. While the 3D-HST fields cover a relatively large skyarea, most regions are restricted to single-PA observations, which limit the ability to distinguish QSOs' broad emission lines from contamination in the host galaxy 2D emission line maps. This limitation also elevates ELG contamination rates when their emission-line regions align with the dispersion axis. To address these challenges, we applied our method to the HST's WISP survey, which includes G102/G141 grism observations across multiple PAs \citep{2010ApJ...723..104A}. Our criteria successfully recovered nearly all known QSOs with sufficient SNR and identified over 30 QSO candidates from the WISP emission-line catalog \citep{2024MNRAS.530..894W}. Notably, follow-up observations using the Hale Telescope are underway to analyze these candidates' properties (Pang et al., in prep).

Several tools beyond Grizli exist for processing slitless spectroscopy, such as the threedhst \citep{2016ApJS..225...27M}, Simulation-Based Extraction (SBE) \citep{2017ApJ...846...84P}, and Linear \citep{2018PASP..130c4501R} procedures. The method discussed in this paper is primarily based on Grizli's forward modeling procedure, but its fundamental approach can be applied to other slitless spectroscopic data processing techniques. For instance, the linear combination factors in Grizli's forward modeling are several physical templates, so when the QSO composite template is added, it improves the fitting results for the QSO's slitless spectrum. During the modeling process, we considered additional factors related to the modeling template, such as the redshift fitting range. Other slitless spectroscopy software, such as Linear, treats the modeling process as a linear factor where a small wavelength segment is constant while the others are zero. In this case, the model spectra of QSOs would show broader emission line components than galaxies. However, this modeling approach also requires considering new adjustment variables, such as the wavelength resolution of the model spectrum and the damping coefficient to reduce the high-frequency oscillations. 

The residuals observed in the line core regions of QSOs in Figure \ref{fig: fitting_re} during the single-template modeling primarily arise from two factors. Firstly, the composite QSO template cannot perfectly simulate broad emission-line profiles across all quasars. While 1D spectral fitting commonly employs multi-Gaussian or Gaussian-Hermite components \citep[e.g.,][]{2012MNRAS.423..600S,2018AJ....155..189D,2019ApJS..241...34S}, implementing such methods in slitless 2D spectroscopy remains constrained by SNR limitations and computational efficiency. Secondly, the reference image are more extended compared to the standard PSF even for the Point-like QSOs, introducing the central modeling discrepancies. Although Grizli's linear fitting approach with the QSO template improves the line wing modeling results, these fundamental limitations persist in the current implementations. Our future work will explore more methods to refine the modeling, particularly for AGN studies in two main aspects:

\begin{itemize}
    \item Multi-component Modeling: Grizli uses an object-based method, which assumes a single SED per source to model 2D grism data. This method is insufficient for weaker AGNs due to significant host galaxy contamination. On the other hand, pixel-based methods treat each pixel as an independent entity, allowing for different SEDs at each location. For AGNs, the optimal modeling approach lies between these two methods: the central AGN component should have a single, well-defined SED with a PSF distribution in the image, while the surrounding host galaxy regions may exhibit different SEDs at different spatial positions. An improved approach should combine the strengths of both pixel-based and object-based methods\textemdash restricting the central pixels to the AGN template while allowing more flexibility in the host galaxy’s outer regions. Ideally, this hybrid method could also yield detailed host galaxy information for weaker AGNs, such as emission-line maps.
    \item Enhanced QSO Templates: Some QSOs show discrepancies between their actual spectra and the composite templates used in our models. We aim to enhance our templates by employing principal component analysis (PCA) for QSO spectra instead of relying on single composite models. Additionally, we will introduce flexible Gaussian profiles with variable FWHM for broad emission lines, using nonlinear modeling techniques to better capture spectral diversity.
\end{itemize}

Finally, incorporating external information, such as multi-band imaging morphology, could further improve the QSO identifications. Using the GaLight \citep{2022ascl.soft09011D} or GALFIT(m) \citep{2002AJ....124..266P,2010AJ....139.2097P} procedure, image decomposition could determine the central PSF flux likely produced by the central AGN. However, challenges still remain. Due to the differences in angular resolution and observational wavelengths between SDSS and HST images, some known QSOs with $z<2$ that appear as typical point sources in SDSS images may exhibit significant host galaxy contributions in the HST images. Moreover, the precise PSF construction is essential for accurate image decomposition. Currently, we have experimented with using the PSF models generated by \cite{2014ApJS..214...24S} and by stacking standard stars which cross-matched with the Gaia catalog \citep{2023A&A...674A..39G}. However, the accurately modeling PSF variations across different image locations remains challenging, leading to potential systematic errors in the decomposition process.

\begin{table*}
	\caption{Trade-off of the $\Delta$BIC criteria}
	\label{tab:criteria-trade-off}
	\begin{tabular}{cccc} 
		\hline
		Criteria & Point-Like QSO & Host-Dominated QSO & ELG pollution (DEEP2) \\
		\hline
            {$ratio_{Q}>$3 \& $\Delta$BIC$<$0} & $\sim$90.4\% (19/21) & $\sim$21.4\% (3/14) & $\sim$4.8\% (15/311) \\
            {$ratio_{Q}>$3 \& $\Delta$BIC$<$-5} & $\sim$81.0\% (17/21) & $\sim$14.3\% (2/14) & $\sim$2.3\% (7/311)\\
            {$ratio_{Q}>$3 \& $\Delta$BIC$<$-30} & $\sim$71.4\% (15/21) & $\sim$0.0\% (0/14) & $\sim$0.3\% (1/311)\\
	   \hline
	\end{tabular}
\end{table*}

\section{Summary and Conclusions} \label{sec:summary}
In this paper, we present a method for selecting QSOs based on their H$\alpha$ and H$\beta$ emission lines using the 3D-HST G141 grism data. Our main conclusions are:

\begin{itemize}
    \item [1)]
    We identify sources with ``broad'' emission lines in the 1D spectra from 3D-HST data release. After excluding sources with contamination, we select about 1000 sources with two or more ``broad'' emission lines and over 5,000 with one ``broad'' emission line in the 3D-HST G141 spectra.
    \item [2)]
    Using forward modeling on known QSO and ELG samples, we developed criteria to distinguish QSOs from ELGs. For known QSOs with H$\alpha$ or H$\beta$ line detected in the G141 band, our criteria successfully selected 62.8\% (22/35) of the QSOs. Most of the missing QSOs (11/14) are dominated by host galaxy light, with host galaxy flux exceeding 75\% in the F140W image. Conversely, QSOs with more point-like structures have a higher success rate of approximately 90\% (19/21). We test our criteria on known ELGs from the DEEP2 and ZFIRE surveys, showing a contamination rate of about 4.8\% (15/311) for the DEEP2 sample, while the 11 ZFIRE ELGs are all excluded from selection.
    \item [3)]
    Applying our method to sources with ``broad'' emission lines, we identified 19 QSO candidates with redshifts ranging from 0.12 to 1.58 and F140W magnitudes between 20.3 and 22.5. Further validation revealed that 3 candidates were confirmed in the DESI EDR, 1 was previously identified as an AGN by \citet{2022ApJ...941..191L}, and 12 had X-ray detections from Chandra. The SED fitting with CIGALE indicated that 17 of the 19 candidates exhibit AGN components.
    \item [4)]
    19 QSO candidates cover a larger color range and are slightly redder in optical slope compared to the SDSS QSO sample, though they remain much bluer than ``little red dots''. Through spectral analysis, we estimate that the candidates have black hole masses of $10^{6.9}-10^{8.3} M_{\odot}$ and Eddington ratios of 0.01-0.03. These candidates improve the completeness of the QSO sample at $z=0.8-1.6$. The number density derived from the 3D-HST QSO sample, including our candidates, aligns well with the QSO luminosity function from \cite{2020MNRAS.495.3252S}.
\end{itemize}

Our new QSO candidates enlarge the QSO sample in the 3D-HST deep fields and demonstrate the feasibility of using slitless spectra to construct representative QSO samples. In future work, we plan to apply our method to other HST surveys to explore the improvement of multiple-position-angle observations and utilizing additional slitless spectroscopy processing tools by combining object-based and pixel-based methods to model the AGNs and their host galaxies. This method could also be applied to upcoming space missions--such as Euclid and the Chinese Space Station Telescope (CSST; \citealp{2025ApJ...980..223P})--as an independent 2D slitless spectra-based QSO selection technique, to build larger and more complete QSO samples.

\section*{Acknowledgements}
We thank Zhiwei Pan, Haicheng Feng, and Bing Lv for helpful discussions. We thank the support of the National Key R\&D Program of China (2022YFF0503401) and the National Science Foundation of China (12133001). This work is based on observations taken by the 3D-HST Treasury Program (GO 12177 and 12328) with the NASA/ESA HST, which is operated by the Association of Universities for Research in Astronomy, Inc., under NASA contract NAS5-26555. This research has made use of data obtained from the Chandra Source Catalog, provided by the Chandra X-ray Center (CXC). We thank the DEEP2 collaboration for making available the high-quality data gathered over several years at the CFHT and Keck observatories. The authors would like to express their gratitude to the reviewer Matthew Malkan for his invaluable feedback and insightful suggestions, which have greatly improved the quality of our work.

\section*{Data Availability}
All the material needed to reproduce Figure \ref{fig:emission_line_properties}, \ref{fig:criteria-figure}, \ref{fig: compare_new_dis}, \ref{fig: compare_color}, \ref{fig: compare_slope}, \ref{fig: LF_compare} and Table \ref{tab:candidates_infor} of this publication is available at this site: \url{https://doi.org/10.5281/zenodo.14830728}. Reduced HST data products can be available upon request. Raw data are available through the MAST website.



\bibliographystyle{mnras}
\bibliography{QSO-3D-HST} 




\appendix
\section{photo-$z$ method} \label{app: photo-z}
We employ the Gradient Boosting Decision Tree (GBDT) for machine learning \citep{friedman2001greedy, Schapire2003}, using the labeled redshift of QSOs from SDSS and Milliquas with the ``Q'' label and the available spectra, in total 61 sources, and all previously spectroscopically confirmed sources with broad emission lines, in total 885 sources in the 3D-HST field to train our model.

The training and validation samples were split at a 3:1 ratio. Due to the relatively small number of QSOs, we increased their weight to improve their performance.

Regarding the feature selection, given the limited size of the training set, we combine the similar photometric bands from different fields into single photometric features after correcting for the Galactic extinction. The final photometric features include magnitudes from the ground-based optical photometry (u, g, r, i, z), ground-based near-infrared photometry (J, H, K), and mid-infrared Spitzer  photometry (IRAC1, IRAC2, IRAC3, IRAC4). Missing photometric values were filled with the mean value. Additionally, we include the adjacent colors within the same photometric system as the input features. In total, 21 features are used for the machine learning.

We optimize the hyper-parameters of GBDT models, including the number of boosting rounds (n\_estimators) and the maximum depth of the decision tree (max\_depth), to achieve the best results for the validation samples. We adopt the normalized median absolute deviation of errors ($\sigma_{\rm{NMAD}}$) and the outlier fraction ($f_c$) as the evaluation metrics for the estimation of the redshift in the training/validation datasets. These metrics are defined as follows: 

\begin{gather}
    \sigma_{\mathrm{NMAD}} = 1.48 \times \mathrm{median} \left(\left| \frac{\Delta z- \mathrm{median}(\Delta z)}{1+z} \right|\right) \\
    f_{\mathrm{c}} = \frac{1}{n} \times \mathrm{count}\left(\left| \frac{\Delta z}{1+z} \right| > 0.15\right),
\end{gather}

\noindent where $z$ is the true redshift, $\hat{z}$ is the predicted redshift, $\Delta z = z-\hat{z}$, and $n$ is the total number of sources. 

\begin{figure}
\centering
\includegraphics[width=\columnwidth]{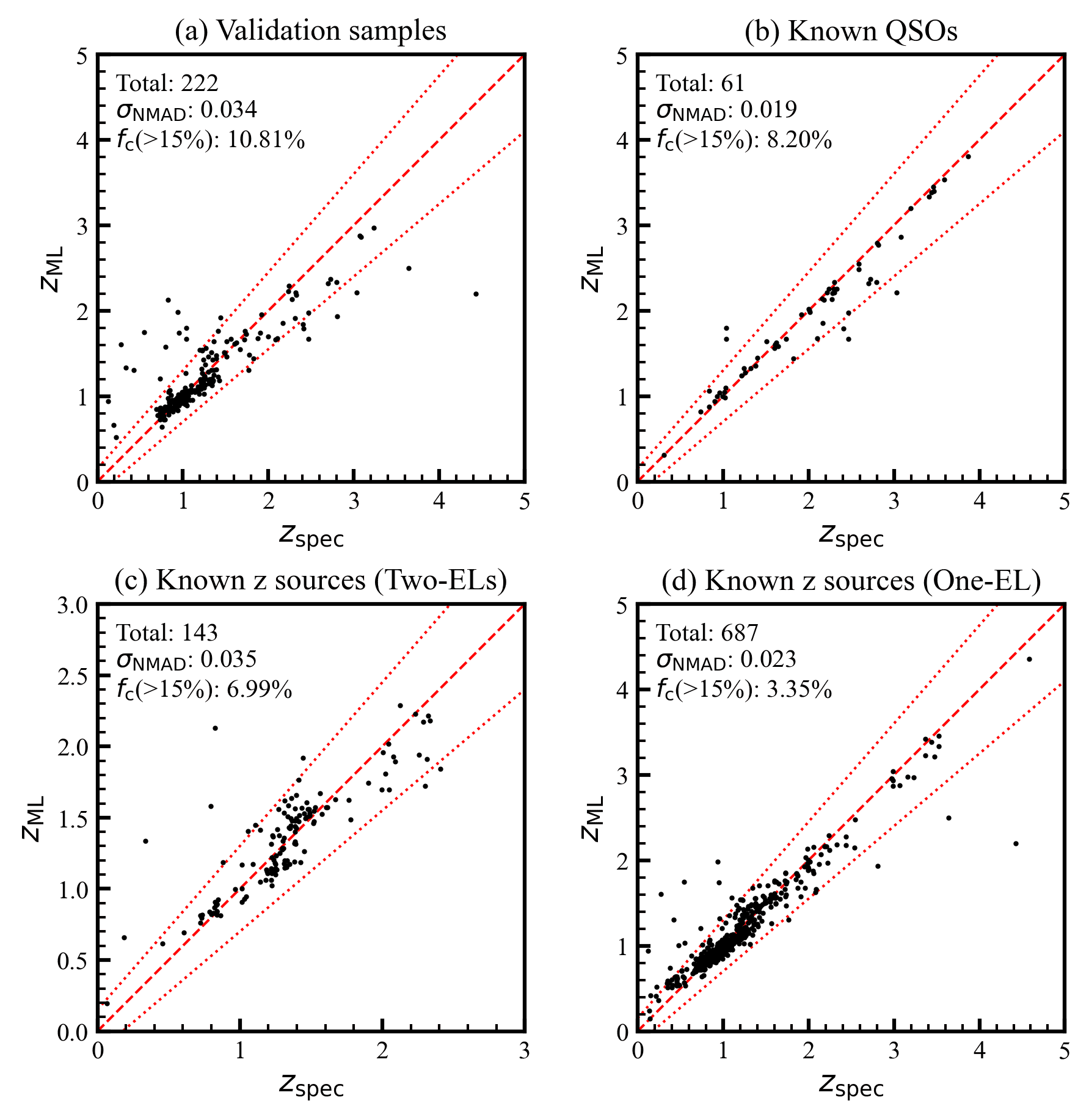}
\caption{Gradient boosting tree photo-$z$ results for different samples. The upper-left and upper-right panels display the results for the validation samples and known QSOs, respectively. The lower-left and lower-right panels present the results for sources with known redshifts in the broad emission line sample selected in Section \ref{subsec:criteria-broadline}. Each panel highlights the sample size, the normalized median absolute deviation of errors ($\sigma_{\rm{NMAD}}$), and the outlier fraction ($f_c$) in the upper left corner.
\label{fig: photo-z train}}
\end{figure}

The final results, as shown in Figure \ref{fig: photo-z train}, represent that both the validation datasets and known QSOs exhibit reasonable photo-$z$ estimates. Around 90\% of the known quasars and validation samples fall within the $0.15\times(1+z_{photo})$ redshift range (between the red dashed lines in the figure), indicating that using such range helps resolve most of the emission line degeneracies.

\section{The G141 Grism Spectra of Candidates} \label{app: can_spec}
The F140W image, G141 2D, and 1D spectra in each dispersion angle of our candidates are plotted in Figure \ref{fig: combine_fig1} and \ref{fig: combine_fig2}.

\begin{figure*}
\centering
\includegraphics[width=0.85\textwidth]{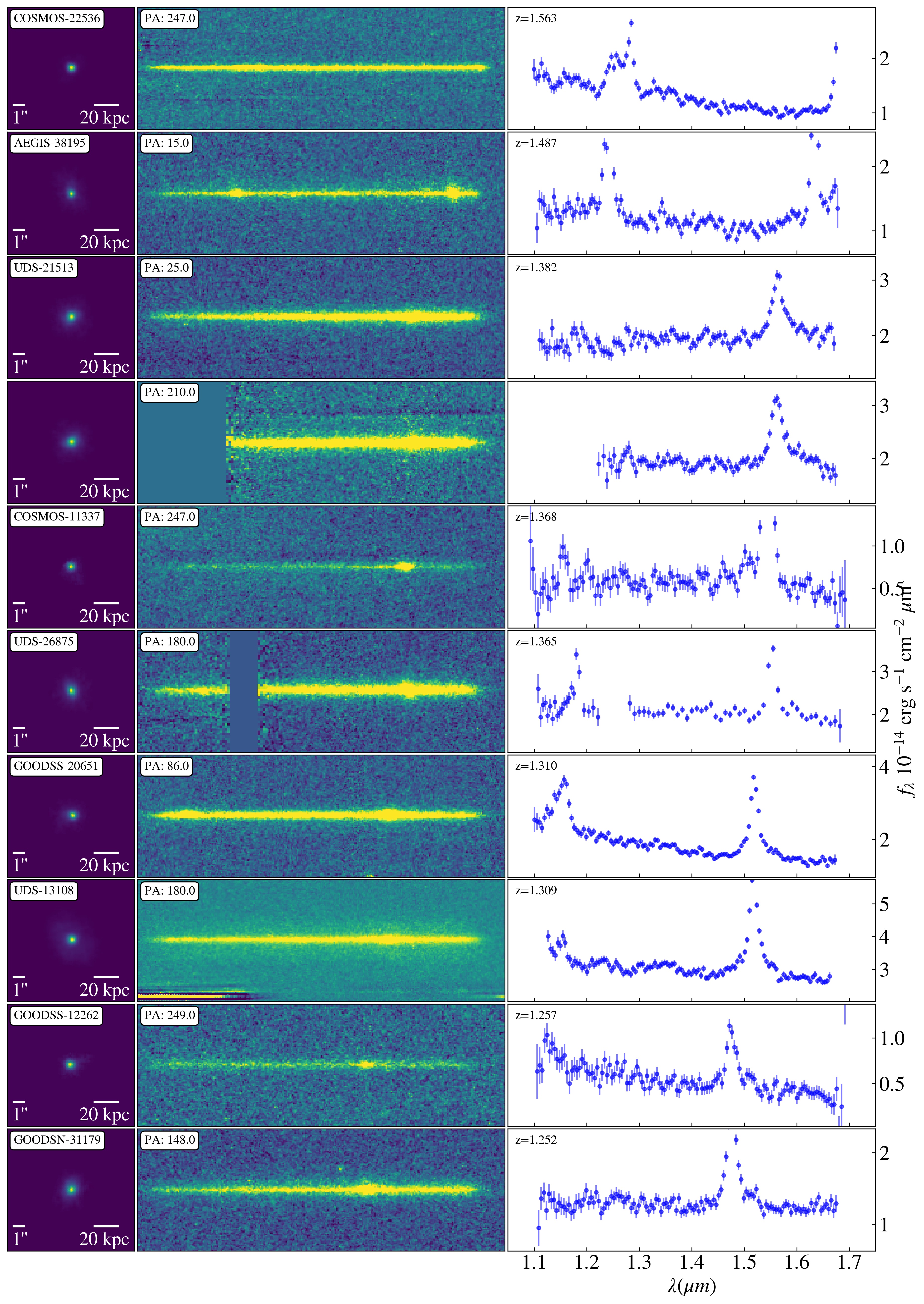}
\caption{The 2D and 1D spectra of candidates, sorted by the redshift. Three panels from left to right show the reference image, 2D G141 spectra, and 1D extracted spectra of each candidate. The note in each panel gives the source name, PA (angle East of North) of the dispersion axis, and redshift of each candidate.
\label{fig: combine_fig1}}
\end{figure*}

\begin{figure*}
\includegraphics[width=0.85\textwidth]{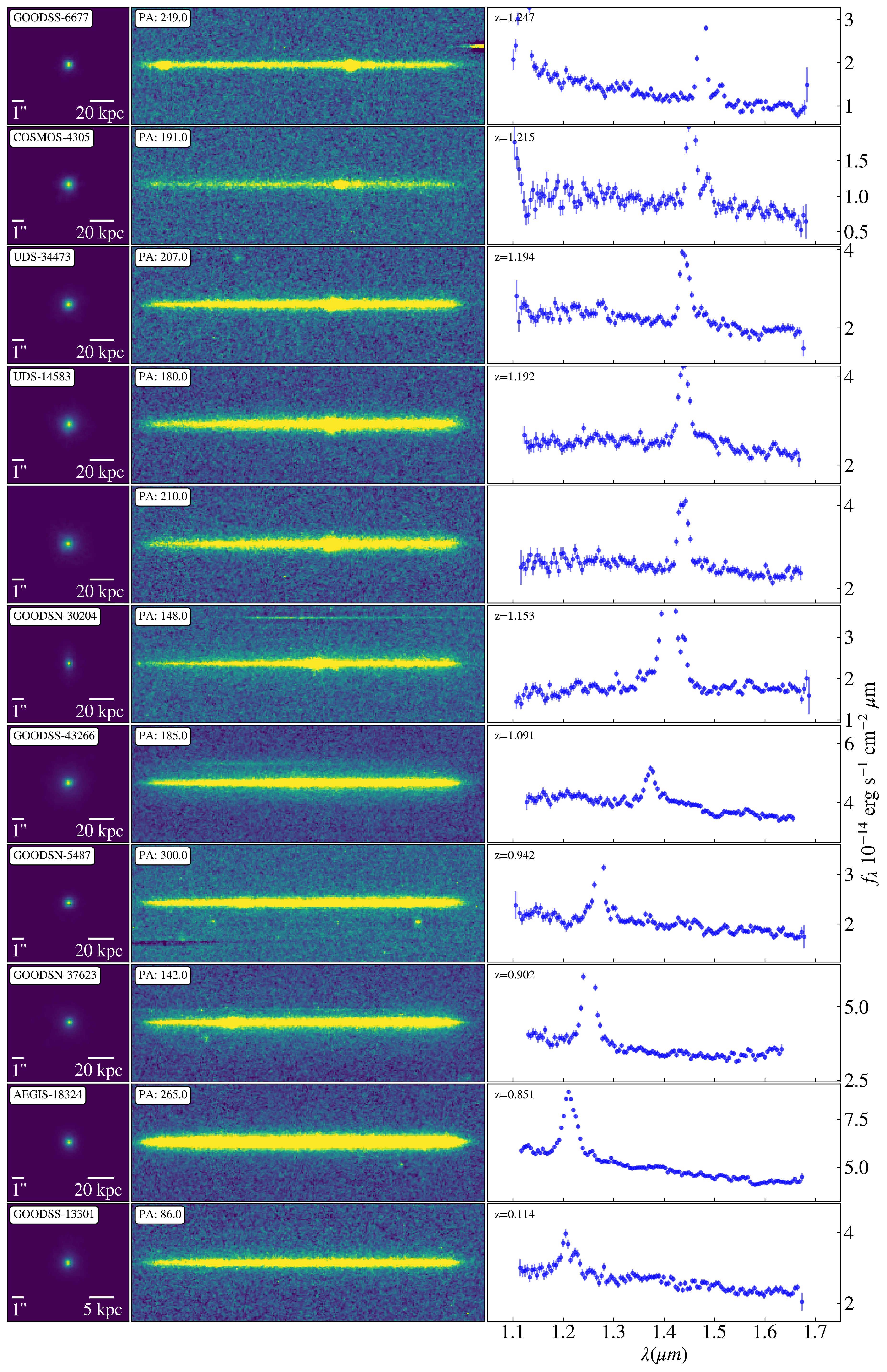}
\caption{Continued image of \ref{fig: combine_fig1}.
\label{fig: combine_fig2}}
\end{figure*}


\bsp	
\label{lastpage}
\end{document}